\documentclass[preprint,showkeys,showpacs,preprintnumbers,amsmath,amssymb]{revtex4}

 \usepackage[]{lineno}

\usepackage[dvips]{graphicx}

\usepackage{bm}        
\usepackage{amssymb, amsmath}
\usepackage{mathrsfs}			
\usepackage{color}

\newcommand{\dims}{{\nu}}		

 \newcommand{\ve}[1]{{\bf #1}}
 \newcommand{\vve}[1]{{\bf #1}}
 \newcommand{\veg}[1]{{\boldsymbol {#1}}}	

\newcommand{\matthree}[9]{\bracket{\begin{array}{ccc}
		#1	&#2	&#3	\\
		#4	&#5	&#6	\\
		#7	&#8	&#9
		\end{array}}}

\newcommand{\vecthree} [3] {\bracket{\begin{array}{c}
		#1 \\
		#2 \\
		#3
		\end{array}} }

\newcommand{\D}{\partial}

\newcommand{\dt}[1]{\frac {d #1} {d t}}

\newcommand{\DI}[1]{\frac {\D #1} {\D x_1}}	
\newcommand{\DIDI}[1]{\frac {\D^2 #1} {\D x_1^2}}

\newcommand{\excl}[1]{{\backslash \hspace{-0.3em} #1}}

\newcommand{\bracket}[1]{\left[#1\right]}
\newcommand{\parenth}[1]{\left(#1\right)}

\newcommand{\ENSO} {{El~Ni\~no\ }}

\DeclareSymbolFont{AMSb}{U}{msb}{m}{n}
\DeclareMathSymbol{\R}{\mathbin}{AMSb}{"52}

\begin{document}

\title{\bf
El Ni\~no Modoki thus far can be mostly predicted \\
more than 10 years ahead of time
}

 \author{X. San Liang}
\email{X.S. Liang, sanliang@courant.nyu.edu;  http://www.ncoads.org/}
\affiliation{Nanjing Institute of Meteorology, Nanjing 210044, China}

\author{Fen Xu}
\affiliation{Nanjing Institute of Meteorology, Nanjing 210044, China}

\author{Yineng Rong}
\affiliation{Nanjing Institute of Meteorology, Nanjing 210044, China}

\date{January 4, 2021}

\vskip 2cm

\begin{abstract}
{
%
%
The 2014-2015 ``Monster''/``Super'' \ENSO failed to be predicted one year
earlier due to the growing importance of a new type of 
El Ni\~no, \ENSO Modoki, which reportedly 
has much lower forecast skill with the 
classical models. In this study, we show that, so far as of today,
this new \ENSO actually 
can be mostly predicted at a lead time of more than 10 years. This is achieved 
through tracing the predictability source with an information flow-based 
causality analysis, which is rigorously established from first principles
in the past decade. We show that the information flowing 
from the solar activity 45 years ago to the sea surface temperature 
results in a causal structure resembling the \ENSO Modoki mode. 
Based on this, a multidimensional system is constructed out of the 
sunspot number series with time delays of 22-50 years. The first 25 
principal components are then taken as the predictors to fulfill the 
prediction, which through information flow-based causal deep learning 
reproduces rather accurately the events 12 years in advance.
%
}
\end{abstract}


\maketitle

\newpage

\vskip 1cm
\noindent
{\Large\bf Revised version submitted to {\it Nature's Scientific Reports}  }

\vskip 1.75cm



\vskip 2.75cm
\noindent
{\bf Submission history:}\\
12/16/2019, Nature, MS\#2019-12-18780, rejected 12/19/2019\\
01/17/2020, PNAS, MS\#2020-01004, rejected 02/11/2020\\
02/19/2020, Nature Communications, MS\#NCOMMS-20-06318-T, rejected 02/21/2020\\
03/06/2020, Science Advances, MS\#abb6053, rejected 03/07/2020\\
04/17/2020, Geophys. Res. Lett., MS\#2020GL088370, rejected 04/25/2020\\
08/12/2020, Nature Climate Change, MS\#NCLIM-20082068, rejected 08/15/2020\\
08/15/2020, Nature's Scientific Reports, 
	    MS\#affcf486-cbcb-4658-9c9d-289452a02b30\\
(Alternative title used in the submissions: Attribution in the presence of
hidden climate drivers---Application shows that \ENSO Modoki is predictable
at a lead time of over 10 years)

\newpage

{\bf
%
%
The 2014-2015 ``Monster''/``Super'' \ENSO failed to be predicted one year
earlier due to the growing importance of a new type of 
El Ni\~no, \ENSO Modoki, which reportedly 
has much lower forecast skill with the 
classical models. In this study, we show that, so far as of today,
this new \ENSO actually 
can be mostly predicted at a lead time of more than 10 years. This is achieved 
through tracing the predictability source with an information flow-based 
causality analysis, which is rigorously established from first principles
in the past decade. We show that the information flowing 
from the solar activity 45 years ago to the sea surface temperature 
results in a causal structure resembling the \ENSO Modoki mode. 
Based on this, a multidimensional system is constructed out of the 
sunspot number series with time delays of 22-50 years. The first 25 
principal components are then taken as the predictors to fulfill the 
prediction, which through information flow-based causal deep learning 
reproduces rather accurately the events 12 years in advance.
%
}




\vskip 0.25cm



\vskip 0.5cm
\ENSO prediction has become a routine practice in operational centers 
all over the world because it helps make decisions in many sectors of 
our society such as agriculture, hydrology, health, energy, to name a few.
Recently, however, a number of projections fell off the mark. 
For example, in 2014, the initially projected ``Monster El~Ni\~no'' did
not show up as expected;
in the meantime, almost no model successfully predicted the 2015 
super \ENSO at a 1-year lead time\cite{McPhaden2015}.
Now it has been gradually clear that there are different types of El Ni\~no;
particularly, there exists a new type that warms the Pacific 
mainly in the center, and tends to be more unpredictable 
than the traditional \ENSO with the present coupled
models\cite{Ashok2009}.
This phenomenon, which has caught much attention 
recently\cite{Trenberth2001}\cite{Larkin2005}\cite{Yu2007}\cite{Ashok2007}\cite{Kug2009}
(c.f.~\cite{Fu1986} for an earlier account), has been termed
\ENSO Modoki\cite{Ashok2007}, Central-Pacific (CP) type
\ENSO\cite{Yu2007}, Date Line \ENSO\cite{Larkin2005}, 
Warm Pool \ENSO\cite{Kug2009}, etc.; see \cite{Wang2017} for a review. 
\ENSO Modoki has left climate imprints which are distinctly different from 
that caused by the canonical El Ni\~no. For example, during \ENSO Modoki episodes, 
the western coast region of the United States suffers from severe drought,
whereas during canonical \ENSO periods it is usually wet\cite{Behera2018}; 
in the western Pacific, opposite impacts have also been realized for 
the tropical cyclone activities during the two types of \ENSO periods.
\ENSO Modoki can reduce more effectively the
Indian monsoon rainfall, exerting influences on the rainfall in
Australia and southern China in a way different from that canonical \ENSO
does; also notably, during the 2009 \ENSO Modoki, a stationary anticyclone
is induced outside Antarctica, causing the melting the ice\cite{McPhaden2015}.

\ENSO Modoki prediction is difficult because
its generating mechanisms are still largely unknown.
Without a correct and complete attribution, the dynamical models may not
have all the adequate dynamics embedded (e.g., \cite{Barnston2005}).
Though recently it is reported to correspond to one of the two most unstable 
modes\cite{Xie2018} for the Zebiak-Cane model\cite{Zebiak1987},
the excitation of the mode, if true, remains elusive. 
%
%
It has been argued that
the central Pacific warming is due to ocean advection\cite{Kug2009}, 
wintertime midlatitude atmospheric variations\cite{Yu2010}, 
wind-induced thermocline variations\cite{Ashok2007}, westerly wind
bursts\cite{Chen2015}, to name a few. 
But these proposed mechanisms are yet to be verified. 	
One approach to verification is to assess the importance of each factor through
sensitivity experiments with numerical models. The difficulty here is,
a numerical climate system may involve many components with parameters 
yet to be determined empirically, let alone the highly nonlinear components,
say, the fluid earth subsystem, may be intrinsically 
uncertain.
Besides, for a phenomenon with physics largely unknown, 
whether the model setup is dynamically consistent is always a concern.
%
For \ENSO Modoki, it has been reported that the present coupled models tend 
to have lower forecast skill for \ENSO Modoki than that for canonical
\ENSO\cite{Ashok2009}, 
probably due to the fact that, during \ENSO Modoki,
the atmosphere and ocean fail to connect.
For all the above, without a correct attribution, 
it is not surprising to see that the 2014-15 forecasts fell off 
the mark--Essentially no model predicted the 2015 \ENSO at 
a one-year lead time (cf.~\cite{Tang2018}).


Rather than based on dynamical models, an alternative approach relies 
heavily on observations and gradually becomes popular as data accumulate.
This practice, commonly known as statistical forecast in climate
science (e.g., \cite{vonStorch2001})
actually appears in a similar fashion in many other disciplines, 
and now has evolved into a new science, namely, data science. 
Data science is growing rapidly during the past few years,
one important topic being predictability.
Recently it has been found that predictability can 
be transferred between dynamical events, and the transfer of 
predictability can be rigorously derived from first principles 
in physics\cite{LK2005}\cite{Liang2016}\cite{Liang2018}. 
The finding is expected to help unravel the origin(s) 
of predictability for \ENSO Modoki, and hence guide its prediction.
In this study, we introduce 
this systematically developed theory, which was 
in fact born from atmosphere-ocean science, 
and, after numerous experiments, show that
\ENSO Modoki actually can be mostly predicted 
at a lead time of 13 years or over, in contrast 
to the canonical El Ni\~no forecasts.
We emphasize that it is {NOT} our intention to do dynamical attributions
here. 
We just present a fact on accomplished predictions.
There is still a long way to go in pursuit of the dynamical processes
underlying the new type of El~Ni\~no.

\vskip 0.5cm
{\bf Seeking for predictor(s) for \ENSO Modoki with information flow
analysis.}
The major research methodology is based on a recently rigorously developed 
theory of information flow, a fundamental notion in physics 
which has applications in a wide variety of 
scientific disciplines (see \cite{Liang2016} and references therein). 
Its importance lies beyond the literal meaning in that it 
implies causation, transfer of predictability, among others.
Recently, it has just been found to be derivable from first principles
(rather than axiomatically proposed),  
initially motivated by the predictability study in atmosphere-ocean
science\cite{LK2005}. 
Since its birth it has been validated with many benchmark dynamical systems
(cf.~\cite{Liang2016}), and has been successfully applied to different 
disciplines such as earth system science, neuroscience, quantitative finance,
etc. (cf. \cite{Vannitsem2019}, \cite{Dionisis2019}, \cite{Stips2016}).
Refer to the Method Section for a brief introduction.

We now apply the information flow analysis to find 
where the predictability of \ENSO Modoki is from. 
{Based on this it is
henceforth possible to select covariate(s) for the prediction.
Covariate selection, also known as sparsity identification, 
is performed before parameter estimation. It arises in all kinds of
engineering applications such as helicopter control\cite{Johnson2005}
and other mechanical system modeling\cite{Bollt2017},
as well as in climate science.
}
We particularly need to use the formula (\ref{eq:T21_hat}) or
its two-dimensional (2D) version (\ref{eq:T21_2d}) as shown in 
the Method Section. 
The data used are referred to the Data Availability Section.

At the first step we sort of use brute force: 
compute the information flows from the index series of some prospect 
driver to the sea surface temperature (SST) at each grid point of 
the tropical Pacific. Because of its quantitative nature, the resulting
distribution of information flow will form a structure which we will refer
to ``causal structure'' henceforth.
While it may be within expectation that information flow may exist
(different from zero) for 
the SST at some grid points, it is not coincidental if the flows 
at all grid points organize into a certain pattern/structure.
(This is similar to how teleconnection patterns 
are identified\cite{Wallace1981}.)
We then check whether the resulting causal structure resembles
the \ENSO Modoki pattern, say, the pattern in the Fig.~2b of Ashok et al.
(2007)\cite{Ashok2007}.
We have tried many indices (e.g. those of 
Indian Ocean Dipole, Pacific Decadal Oscillation, North Atlantic, Atlantic
Multi-decadal Oscillation,
to name but a few), and have found that the series of sunspot numbers (SSN)
is the very one. In the following we present the algorithm and
results.



We first use the 2D version of Eq.~(\ref{eq:T21_hat})
(cf.~(\ref{eq:T21_2d})) to do a rough estimate for
        the information flow from SSN to the SST at each grid point,
	given the time series as described in the preceding section. 
	This is the practice how teleconnection patterns are identified using
	correlation analysis, but here the computed is information flow. 
	(Different from the symmetric correlation, here the information 
	flow in the opposite direction is by computation insignificant, 
	just as expected.)
We then form delayed time series for SSN, given time delays $n=1,2,...$, 
	and repeat the above step. For convenience, denote an SSN series with
delay $n$ by SSN($-n$).
{
We find that the resulting information flow is not significant at a 90\%
confidence level, or has a structure bearing no resemblance to the desired pattern,
until $n$ approaches 45 years; see Supplementary Figure~S1, 
for a number of examples. }
Here the data we are using are the SST from 01/1980 through 12/2017 (SST are
most reliable after 1979), and the SSN data are correspondingly from
01/1935 through 12/1972. 
Note that using the time delayed coordinates we can reconstruct a dynamical
system which is topologically equivalent to the one that originally
generates the time series, thanks to Takens' embedding theorem\cite{Takens1980}. Here, based on the above rough estimates, we choose to form a
dynamical system with components corresponding to the SST series and 
the time-delayed series SSN($-n$) with $n$ ranging from 22-50 years.
Note one is advised not to choose two delays too close.
SSN is dominated by low-frequency processes; two close series thus-formed
may not be independent enough to span a subspace, and hence 
may lead to singularity.
Here we choose SSN series at delays of 22-50 years every 5 years 
or 60 steps. We have tried many other sampling intervals and found this 
is the best. This makes sense: An interval of 5 years put the two series
approximately at the opposite phases in a solar cycle.
From the thus-generated multivariate series, we then compute the information flow 
from SSN($-45$ years) to the Pacific SST, using formula (\ref{eq:T21_hat}). 

{
A note on the embedding coordinates. In general, it is impossible to 
reconstruct exactly the original dynamical system from the embedding
coordinates; it is just a topologically equivalent one. That also implies 
that the reconstruction is not unique. Other choices with delayed time 
series may serve the purpose as well, provided that they are independent 
of other coordinates; see \cite{Abarbanel1996} for empirical methods 
for time delay choosing. Fortunately, this does not make a problem for our
causal inference, thanks to a property of the information flow {\em ab
initio}, which asserts that 
	\begin{itemize}
        \item[] The information flow between two coordinates of
	a system stays invariant upon arbitrary nonlinear transformation 
	of the remaining coordinates.
	\end{itemize}
More details are referred to the Method Section.
This remarkable property implies that the embedding coordinates do not
matter in evaluating the information flow from SSN to SST; what matters is
the number of the coordinates, i.e., the dimension, of the system, 
which makes a topological invariant. There exist empirical ways to
determine the dimension of a system, e.g., those as shown in 
\cite{Abarbanel1996}. Here we choose by adding new coordinates and examining 
the information flow; if it does not change any more, the process stops and
the dimension is hence determined. For this problem, as shown above, the 
SSN series at delays of 22-50 years every 5 years 
are chosen. This results in 6 auxiliary coordinates, which together with
the SST and SSN(-45 years) series make a system of dimension 8.
}

\begin{figure}[h]
	\begin{center}
	\includegraphics[angle=0,width=0.9\textwidth] 
	{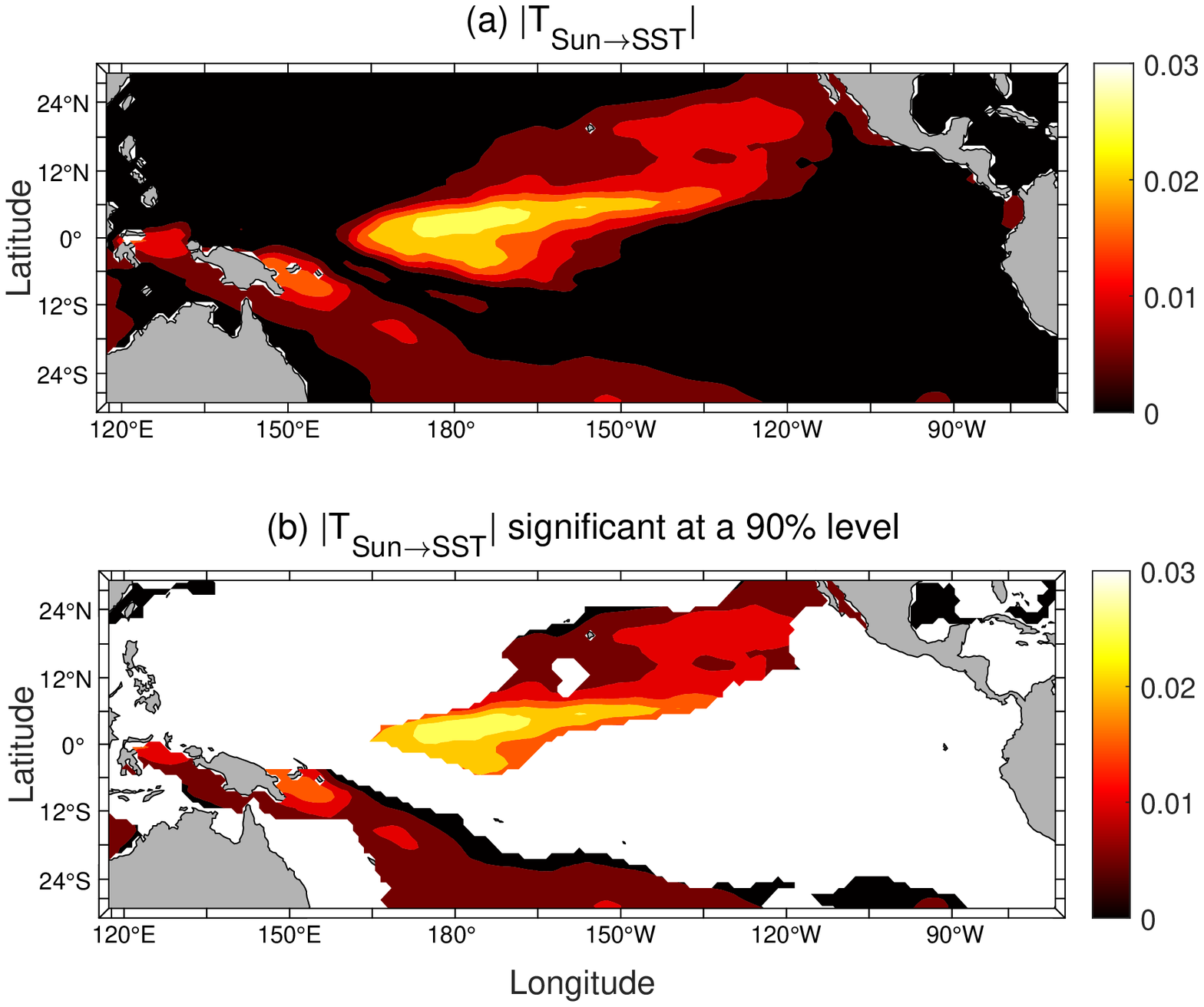}
	\caption{(a) The absolute information flow from the delayed 
		(by 45 years) series of sunspot numbers 
		to those of the sea surface temperature in the Pacific Ocean 
		(in nats/month). The pattern resembles very much
		the \ENSO Modoki mode as shown in the 
		Fig.~2b of Ashok et al. (2007)\cite{Ashok2007}.
		Here information flow means the transfer of predictability
		from one series to another. It is computed with
		Eq.~(\ref{eq:T21_hat}) in section~2.
		(b) As (a), but only the information flow significant 
		at a 90\% confidence level is shown.
		\protect{\label{fig:causal_pattern}}}
	\end{center}
	\end{figure}

The spatial distribution of the absolute information flow, or ``causal
structure'' as will be referred to henceforth, from 
SSN($-45$~years) to the Pacific SST is shown in
Fig.~\ref{fig:causal_pattern}. 
Remarkably,
the causal structure in Fig.~\ref{fig:causal_pattern}a
is very similar to 
the \ENSO Modoki pattern, as given by, say, Ashok et al.\cite{Ashok2007}.
(Please see their Fig.~2b.)
That is to say, the information flow from SSN to the Pacific SST does form 
a spatially coherent structure resembling to \ENSO Modoki.  
The computed information flow is significant at a 90\% confidence level, 
as shown in Fig.~\ref{fig:causal_pattern}b.

Comparing to the rough estimate in the bivariate case (see Supplementary
Figure~S1d),
Fig.~\ref{fig:causal_pattern}a has a horseshoe structure pronounced
in the upper part, plus a weak branch in the Southern Hemisphere. This is
just as the \ENSO Modoki structure as obtained by Ashok et
al.\cite{Ashok2007} (see their Fig.~2b). This indicates that the bivariate
information flow analysis provides a good initial guess, but the
result is yet to be rectified due to the incorrect dimensionality.

The information flows with other time delays have also been computed. Shown
in Supplementary Figure~S2, are a number of examples. As can be seen, except those
around 45 years (about 44-46 years, to be precise), other delays either 
yield insignificant information flows or do not result in the 
desired causal pattern.

Since we are about to predict, it is desirable to perform the causality
analysis in a forecast fashion, i.e., to pretend that the EMI over the
forecast period be unavailable. We have computed the information flow with
different years of availability. Shown in Fig.~S3 is the information flow
as that in Fig.~\ref{fig:causal_pattern}, but with SST data available until
the year of 2005. Obviously, the causal pattern is still there and
significant, and, remarkably, it appears even more enhanced, probably 
due to the frequent occurrences of the event during that period.

As information
flow tells the transfer of predictability, SSN forms a natural predictor
for \ENSO Modoki. This will be further confirmed in the next section.


\vskip 0.5cm
{\bf Projection of \ENSO Modoki.}
The remarkable causal pattern in Fig.~\ref{fig:causal_pattern}
implies that it may be possible to make projections of \ENSO Modoki 
many years ahead of time.	
{
Based on it we hence conjecture that the lagged SSN series be 
the desired covariate.}
Of course, due to the short observation
period, the data needed for parametric estimation are rather limited--
This is always a problem for climate projection.

\begin{figure}[h]
	\begin{center}
	\includegraphics[angle=0,width=0.75\textwidth, height=0.8\textwidth]
		{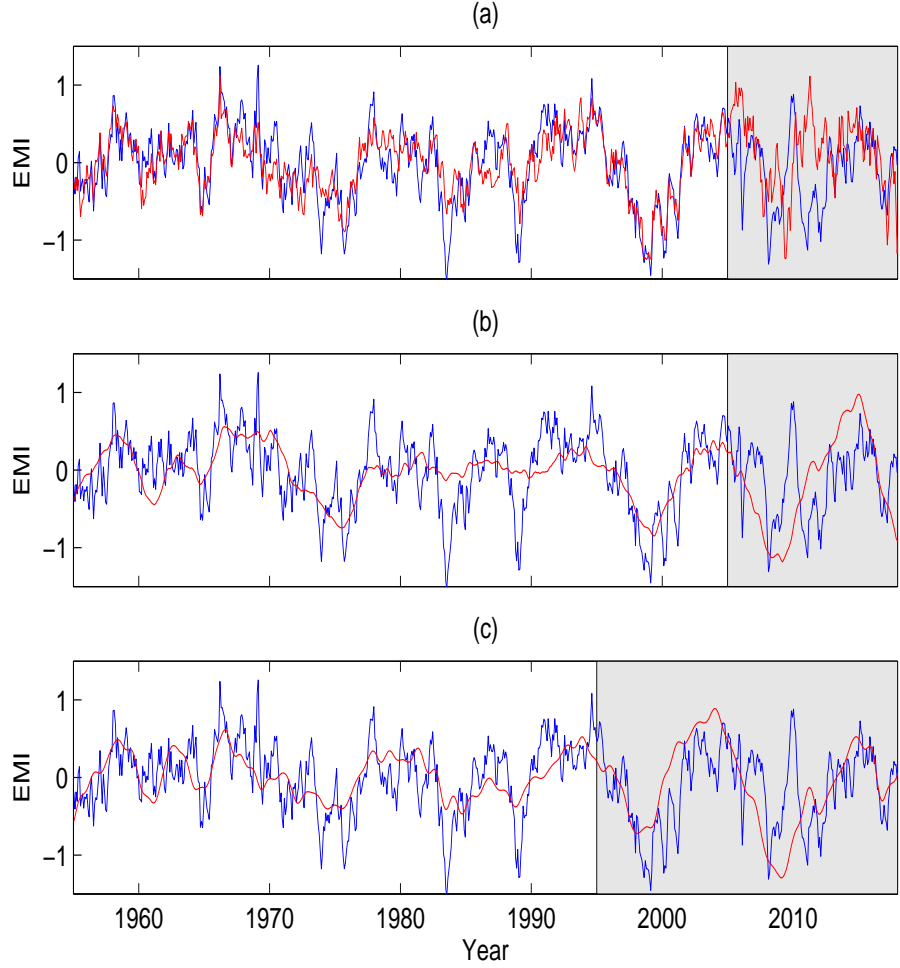} 
	\caption{Projection of the \ENSO Modoki index based solely on 
		the sunspot numbers 22-50 years ago with
		a simple linear regression model. 
		(a) The SSN series is bandpass filtered with the annual
		signals and the signals longer than 11 years removed.
		(b) Only 25 first principal components are used
		as inputs.
		(c) The SSN series is bandpass filtered as in (a), and only
		    the 25 first principal components are used as inputs. 
		The shaded period is for prediction (note in (c) the
		period for prediction is 1995-), and the 
		predicted index is in red.
		\protect{\label{fig:prediction}}}
	\end{center}
	\end{figure}

Let us start with a simple linear regression model for the prediction of
\ENSO Modoki Index (EMI) from SSN. Build the model using the EMI data from 
{1975-2005 ($30\times12=600$ equations)},
leaving the remaining years after 2005 for prediction. For each EMI, 
the model inputs are the SSNs at lead times of 50 years through 22 years
(336 in total), guided by the causality analysis in the preceding section. 
For best result, the annual signals and the signals longer than 11 years
are filtered from the SSN series. 
{The filtering is fulfilled through wavelet analysis with the
orthonormal spline wavelets built in \cite{LA2007}. Since it is required
that the series length be a power of 2, we choose a time range for the
monthly SSN series from May 1847 to December 2017, depending on the 
availability of 
the data when this research was initialized. This totals 170 years and 8
months (170.67 years), or $2^{11}=2048$ time steps. 
The upper bound scale level for the wavelet analysis is set $8$,
which gives a lower period of $2^{-8} \times 170.67 = 0.67$ years;
the lower bound scale level is chosen to be 4, resulting an upper period of
$2^{-4} \times 170.67 = 10.67$ years. In doing this the seasonal cycle and
the interdecadal variabilities are effectively removed (Supplementary
Figure S4). 
}
With the pretreated SSN series the prediction is launched, and 
the result is shown in Fig.~\ref{fig:prediction}a.

We would not say that the projection in Fig.~\ref{fig:prediction}a
with such a simple linear model is successful,
{
but the 2015/2016 event is clearly seen. 
The strong 2009 event is not correct; it has a phase error. 
The general trend between 2005-2018 seems to be fine. Beyond 2018 it
becomes off the mark.}

{
It would be of some use to see how the simple linear prediction may 
depend on some factors, though it cannot be said successful and hence a
quantitative investigation does not make sense. We hope to learn some
experience for the machine learning to be performed soon.
First, we check the effect of filtering.
As shown in Supplementary Figure~S5, 
without filtering, the result is much noisy, the 2015/2016 event 
is too strong, and the strong 2009/2010 event completely disappears; 
with low-pass filtering (only the signals below a year are filtered), 
the result is similar, only with the noise reduced and the 2015/2016 event
better.

Second, we perform an empirical orthogonal function (EOF) analysis 
for the 336 time-delayed SSN series, which forms a column vector
with 336 entries at each time step:
   [${\rm SSN}(n-600), {\rm SSN}(n-599), ..., {\rm SSN}(n-265)$]$^T$.
The variance for each EOF mode is plotted in Supplementary Figure~S6. From it
the first 8, 25, and 50 principal components (PCs) approximately 
account for 84\%, 89\%, and 92\% of the total variances, respectively. 
Some examples of the EOF modes are plotted in Supplementary Figure~S7. 
Now use the first 25 PCs as inputs and repeat the above process to make the
projection. The result is shown in Fig.~\ref{fig:prediction}b.
Obviously, during the 12 years of prediction from 2005 to 2017, 
except for the strong 2009/2010 event, the others are generally fine.
Projections with other numbers of PCs have also been conducted. 
Shown in Supplementary Figure~S8 are examples with 8 and 50 PCs,
respectively. 
It is particularly interesting to see that, except for the 2009/2010 event,
the case with 8 PCs already captures the major trend of the EMI evolution.
 Recall that, in performing the information flow analysis, 
 we have found that the reconstructed dynamical system approximately 
 has a dimension of 8. Here the prediction agrees with the dimension
 inference.
}

\begin{figure}[h]
	\begin{center}
	\includegraphics[angle=0,width=0.8\textwidth] {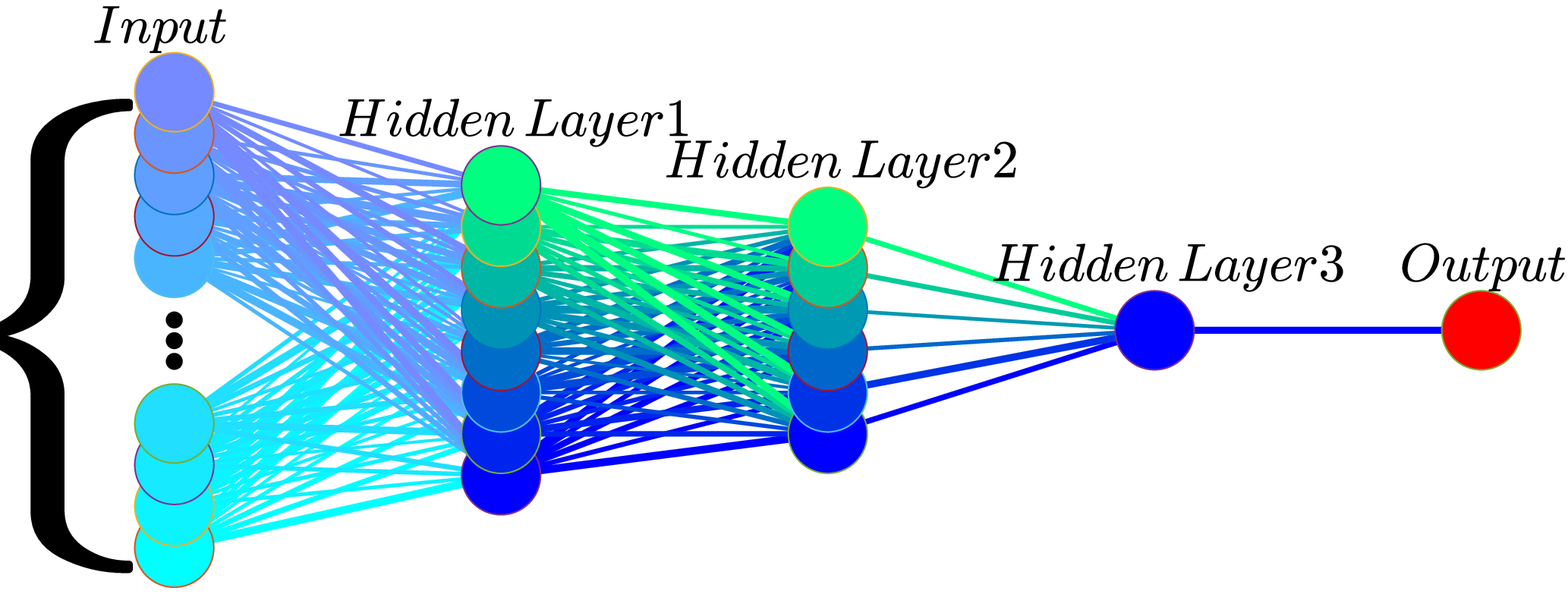}
	\caption{The neural network architecture used in this study, 
	which is made of an input layer, an output layer, 
	and three hidden layers.
	The symbols are the same as those in (\ref{eq:tansig}).
	The matrices/vectors $\ve w_i$ and $\ve b_i$ ($i=1,2,3,4$)
	are obtained through training.
	To predict the \ENSO Modoki Index (EMI) at month/step $n$, 
	${\rm EMI}(n)$, find the first 25 principal components of
	[${\rm SSN}(n-600), {\rm SSN}(n-599), ..., {\rm SSN}(n-265)$]$^T$, 
	a column vector with 336 entries,
	and form the input vector $\ve x$.
	The final output $y$ is the predicted ${\rm EMI}(n)$. 
	Iterating on $n$, we can get a prediction of EMI for any target
	time interval.
		\protect{\label{fig:BP_architech}}}
	\end{center}
	\end{figure}

Because of the encouraging linear model result, it is desirable 
to achieve a better projection using more sophisticated tools 
in deep learning, e.g., the back propagation (BP) neural network 
algorithm\cite{Goodfellow2016}. 
(At the preparation of the original version of this manuscript,
we noticed that recently there have been applications of 
other machine learning methods to \ENSO forecasts, e.g., \cite{Ham2019}.)
Three hidden layers are used for the BP neural network forecast.
Again, a major issue here is the short
observational period which may prevent from an appropriate training with
many weights. 
{We hence use only 8, 6, and 1 neurons for the
first, second, and third hidden layers, respectively, 
as schematized in Fig.~\ref{fig:BP_architech}.} This architecture will be
justified soon.
To predict an EMI at step (month) $n$, written $y$, a $336\times1$ 
vector $\ve x$ is formed with SSN data at steps (months) 
	  $n-600, n-599, ..., n-265$,
the same as the case with lead times of 50-22 years
for the above simple linear regression model. 
But even with the modest number of parameters, the data are still not
enough. To maximize the use of the very limited data, 
we choose the first 25 principal components (PCs) that account for 90\% 
of the total variance. 
We hence train the model with these 25 inputs (rather than 336 inputs);
that is to say, $\ve x$ now is a $25\times 1$ vector, which
greatly reduces the number of parameters to train.
(See below in Eq.~(\ref{eq:tansig}):
The $\ve w_1$ is reduced from a $8\times336$ matrix to a $8\times25$
matrix.)
Once this is done, $\ve x$ is input 
into the following equation to arrive at the prediction of $y$:
	\begin{eqnarray}		\label{eq:tansig}
	y = {\vve w_4}\ {\rm tansig} ({\vve w_3}\ 
		{\rm tansig} (\vve w_2\ {\rm tansig}(\vve w_1\ \ve x  
		+ \ve b_1) + \ve b_2)  +  \ve b_3)  +  \ve b_4,
	\end{eqnarray}
where the matrices/vectors of parameters 
 $\vve w_1$ ($8\times25$ matrix),
 $\vve w_2$ ($6\times8$ matrix),
 $\vve w_3$ ($1\times6$ row vector),
 $\vve w_4$ ($1\times1$ matrix---a scalar),
 $\vve b_1$ ($8\times1$ vector),
 $\vve b_2$ ($6\times1$ vector), 
 $b_3$ (scalar) and $b_4$ (scalar)
are obtained through training. 
An illustrative explanation is seen in Fig.~\ref{fig:BP_architech}.
{As has been argued (e.g.~\cite{Heaton2008}) that
the number of hidden neurons should be kept fewer than
2/3 the size of the input and output layers, i.e., 
$\frac23\times (25+1)=17$ here. With this architecture we have 
15 neurons in total, meeting the requirement.}

%
%

To predict the EMI at month/step $n$, 
	${\rm EMI}(n)$, organize ${\rm SSN}(n-600), {\rm SSN}(n-599), 
	..., {\rm SSN}(n-265)$, 
	that is, the SSNs lagged by 50 through 22 years,
into a $336\times1$ vector. 
Project it onto the first 25 EOFs as obtained above,
and obtain $\ve x$, a vector with 25 entries.
Input $\ve x$ into Eq.~(\ref{eq:tansig}). The final output $y$ is the
predicted ${\rm EMI}(n)$. Iterating on $n$, we can get a prediction of EMI
for any target time interval. 

\begin{figure}[h]
	\begin{center}
	\includegraphics[angle=0,width=0.75\textwidth] {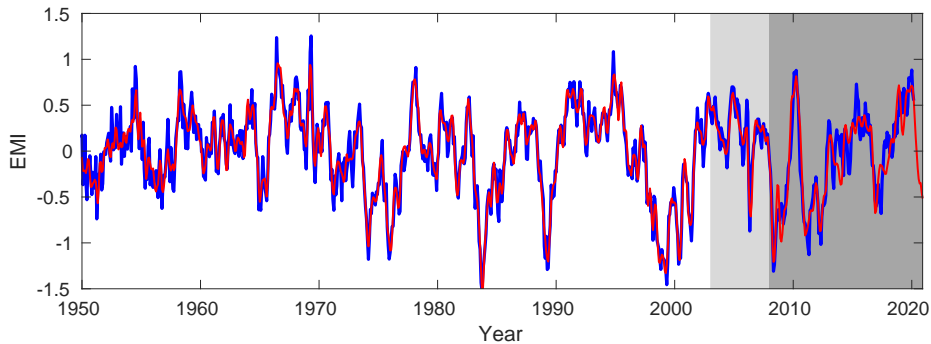}
	\caption{Projection of the \ENSO Modoki index based solely on 
		the sunspot numbers 22-50 years ago with
		the back propagation neural network algorithm.
		The EMI is in blue, while the predicted index is in red.
		Lightly shaded is the period over which the validation set
		is formed, and the period with dark shading is for
		prediction.
		\protect{\label{fig:BP_prediction}}}
	\end{center}
	\end{figure}

{
The 5-year data prior to the starting step of prediction are used
for validation.
In this study, we start off predicting EMI at January 2008.
We hence take the data over the period from January 2003 through December 2007 
to form the validation set, and the data until December 2002 to train 
the model. 10,000 runs have been performed, and we pick the one that
minimizes the mean square error (MSE) over the validation set. This is 
plotted in Fig.~\ref{fig:BP_prediction}. 
Remarkably, the 12-year long index has been forecast with high 
skill (corr. coeff.=0.91).
Particularly, the strong 2009/10 event is well predicted, 
so is the 2019/20 event. The 2014/16 event, which has made the \ENSO forecasts 
off the mark in 2014-15, appears a little weaker but also looks fine by
trend.}

\begin{figure}[h]
	\begin{center}
	\includegraphics[angle=0,width=0.75\textwidth] 
	{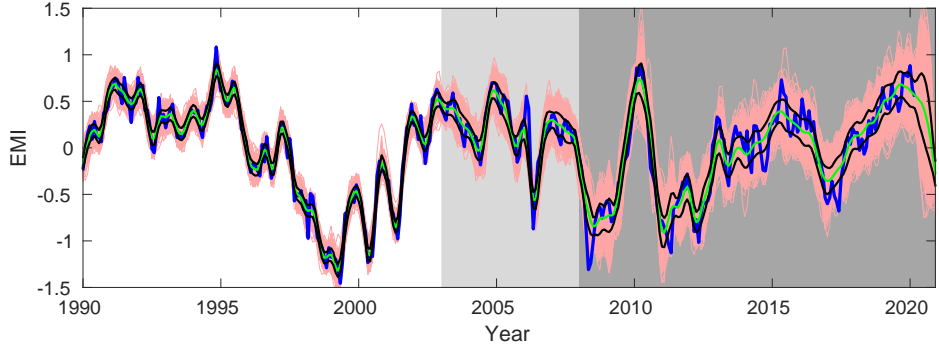}
	\caption{The 1000 predictions showing the spread of the ensemble.
		Overlaid are the observed EMI (blue), the mean of the
		realizations (cyan), and the mean plus and minus the
		standard deviation (black). The light shading 
		marks the period for validation, while the darker shading
		marks the prediction period.
		\protect{\label{fig:BP_uncertain}}}
	\end{center}
	\end{figure}

{
Out of the 10,000 runs we select the top 10\% with lowest MSE over the
validation set (January 2003-December 2007) to form an ensemble of 
predictions for the 12 years of EMI (January 2008 till now). The spread,
the mean, and the standard deviation of the ensemble are shown in
Fig.~\ref{fig:BP_uncertain}. Also overlaid is the observed EMI (blue).
As can be seen, the uncertainty is rather 
limited for a climate projection.}

\begin{figure}[h]
	\begin{center}
	\includegraphics[angle=0,width=0.75\textwidth] {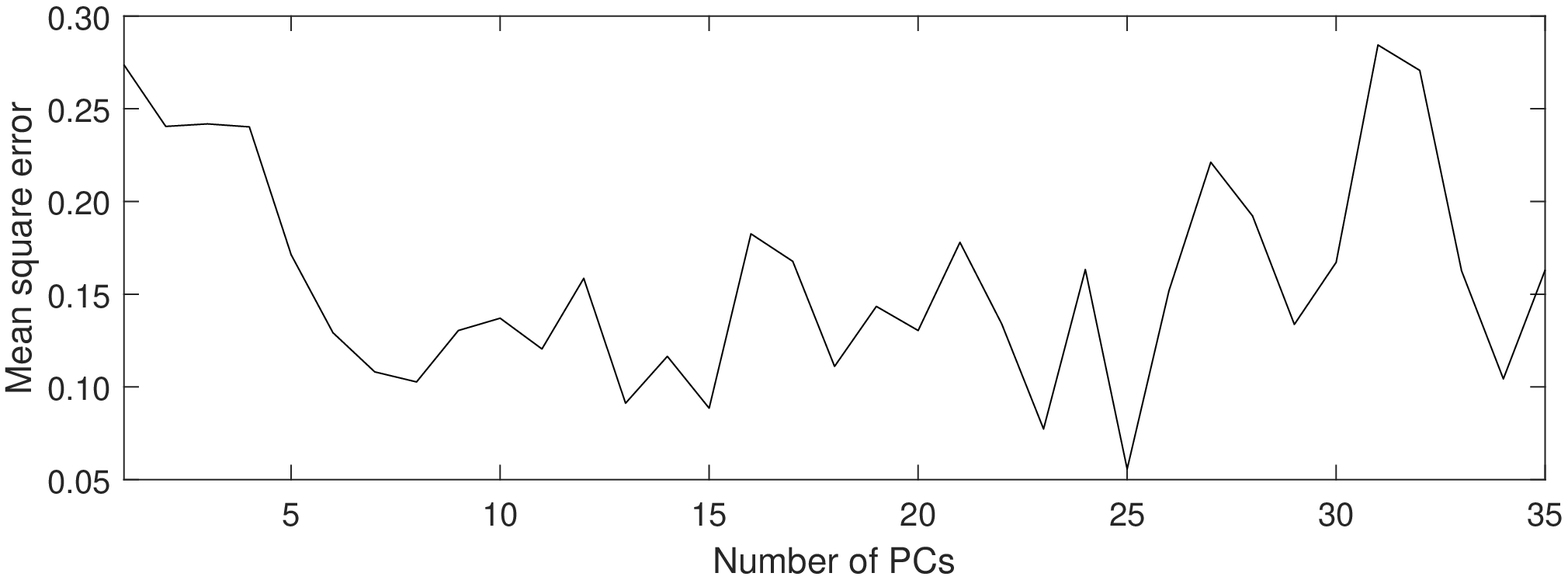}
	\caption{Mean square error for the neural network prediction
		as a function of number of principal components.
		\protect{\label{fig:mse_pc}}}
	\end{center}
	\end{figure}

{
We have examined how the projection performance may vary with the choices 
of PCs, the number of neurons, and the lags.
The PCs of a vector are obtained by projecting it onto the EOF modes
(described above), which possess variances and structures as plotted in 
Supplementary Figures S6 and S7, respectively. 
Shown in Fig.~\ref{fig:mse_pc} is the MSE as a function of 
the number of PCs. Obviously with 25 PCs the MSE is optimized, and that is
what we have chosen for the network architecture. But if one takes a closer
look, the MSE does not decrease much beyond 8. Recall that, 
in performing the causality analysis, we have seen that 
the reconstructed dynamical system approximately has a dimension 8. 
Fig.~\ref{fig:mse_pc} hence provides a verification of the dimension
inference.
}

{
The EOF modes associated with these 8 PCs are plotted in Supplementary
Figure~S7.
As can be seen, except for the 11-year process, among others, 
the 5-year variability is essential. 
This is, again, consistent with what we have
done in forming the embedding coordinates: select the delayed series 
every 5 years.
}

\begin{figure}[h]
	\begin{center}
	\includegraphics[angle=0,width=0.75\textwidth] {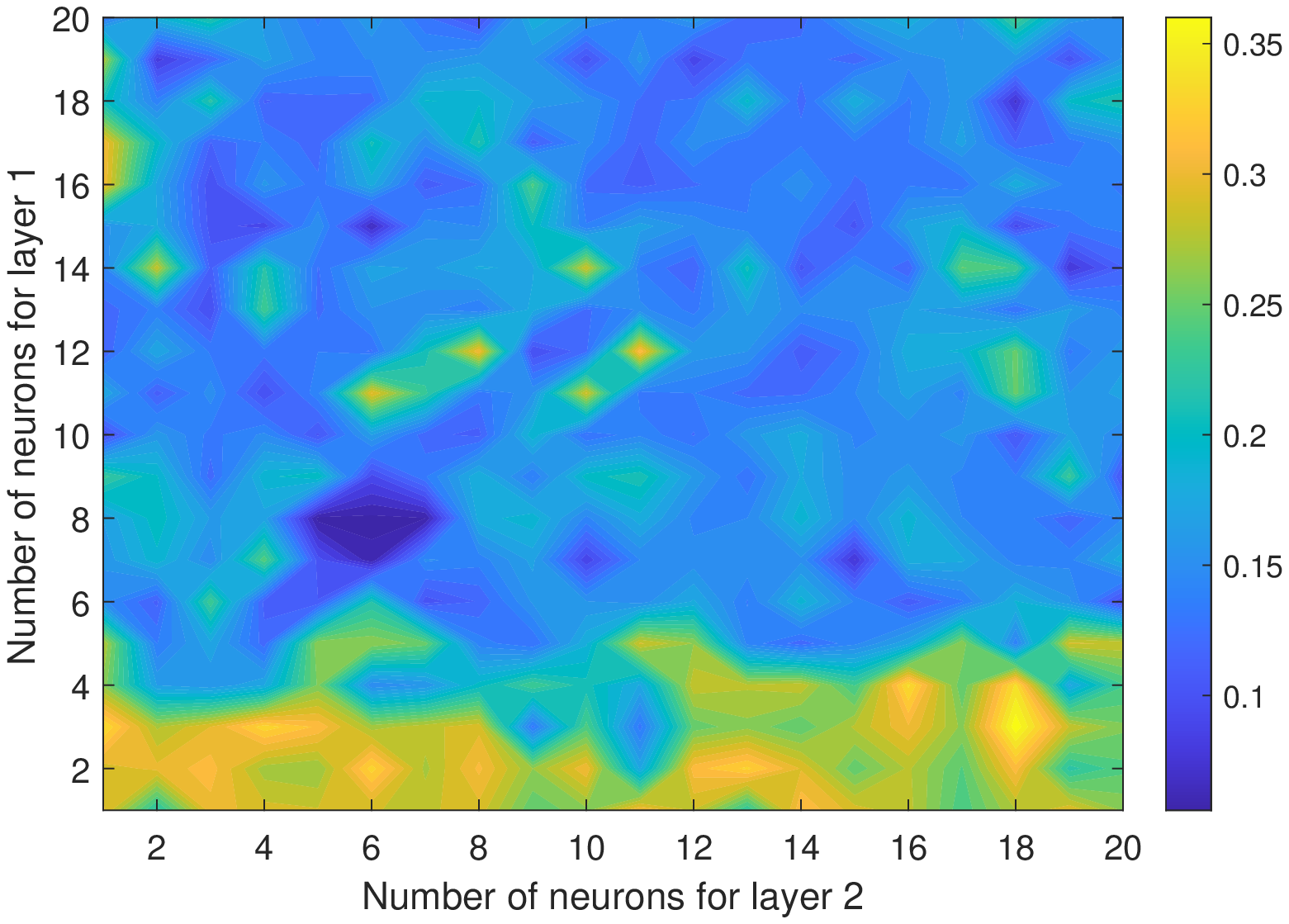}
	\caption{Mean square error as function of number of neurons.
		\protect{\label{fig:mse_neuron}}}
	\end{center}
	\end{figure}

{
The dependence of the model performance on the number of neurons has also
been investigated. We choose 1 neuron for layer 3, leaving the numbers of
layers 1 and 2 for tuning. The result is contoured in 
Fig.~\ref{fig:mse_neuron}. As can be clearly seen,
a minimum of MSE appears at (8,6), i.e., when the numbers of
hidden layer 1 and layer 2 are, respectively, 8 and 6. 
This is what we have chosen to launch the standard prediction.
}

\begin{figure}[h]
	\begin{center}
	\includegraphics[angle=0,width=0.75\textwidth] {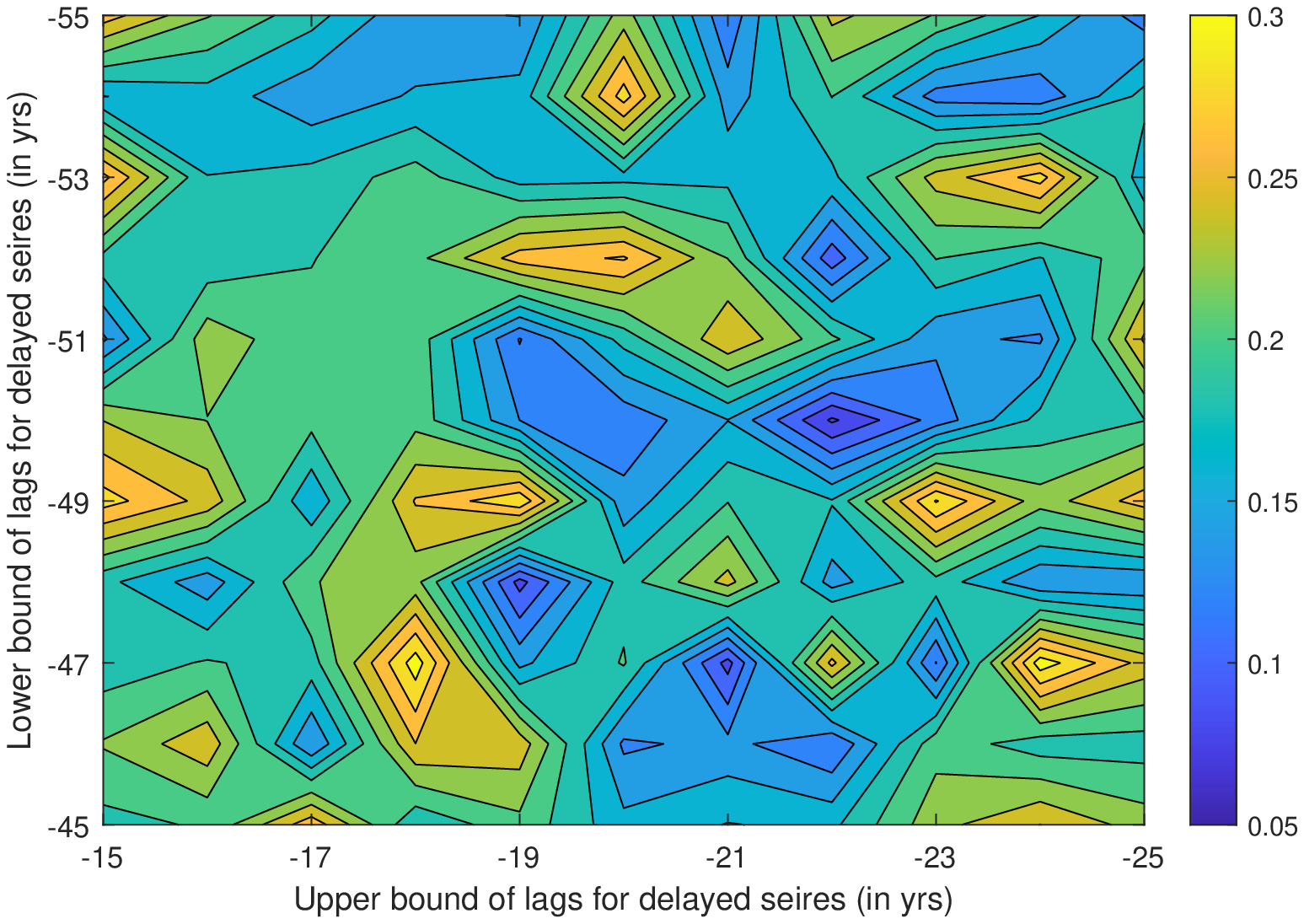}
	\caption{Mean square error as function of lags.
		\protect{\label{fig:mse_lag}}}
	\end{center}
	\end{figure}

{
Also studied is the influence of the lags for the delayed series. 
In Fig.~\ref{fig:mse_lag} the MSE is shown as function of the lower and
upper bounds of the delays (in years). From the figure there are a variety
of local minima, but the one at (22, 50) is the smallest one. That is to
say, the series with delays from 22 to 50 years make the optimal
embedding coordinates, and this is just we are choosing.
}

{
It is known that machine learning suffers from the problem of 
reproducibility\cite{Hutson2018}. But in this study, as we have shown
above, the prediction guided by the information flow analysis
is fairly robust, with uncertainty acceptably small for climate projection.
}
For all that account, at least the \ENSO Modoki events so far
are mostly predictable at a lead time of more than 10 years.

\vskip 0.5cm
{\bf Concluding remarks.}
%
%
%
The rigorously developed theory of information flow in terms of
predictability/uncertainty transfer provides a natural way for 
one to seek for predictor(s) for a dynamical phenomenon. 
In this study, it is found that the delayed causal pattern from SSN to the
Pacific SST resembles very much the \ENSO Modoki mode; the former hence
can be used to predict the latter.
Indeed, as detailed above, with all the observations we have had so far, 
\ENSO Modoki can be essentially predicted based solely on SSN
at a lead time of 12 years or over.
Particularly, the strong event in 2009/10 and the elusive event during
2014-16 have been well predicted (Fig.~\ref{fig:prediction}b).


We, however, do NOT claim that \ENSO Modoki is ultimately driven by solar
activities. It is NOT our intention in this study 
to investigate on the dynamical aspects of this climate mode.
We just present an observational fact on fulfilled predictions.
There is still a very long way to go in unraveling the dynamical 
origin(s) of \ENSO Modoki. 
The success of atmosphere-ocean coupled models during the past
decades confirms that the canonical \ENSO is an intrinsic mode in the
climate system; \ENSO Modoki may be so also.
Nonetheless, despite the long-standing controversy on the role of solar 
activity in climate change (see a review in \cite{Bard2006}), 
there does exist evidence on the lagged response as identified here;
the North Atlantic climate response is such an example\cite{Scaife2013}.

\begin{figure}[h]
	\begin{center}
	\includegraphics[angle=0,width=1\textwidth] {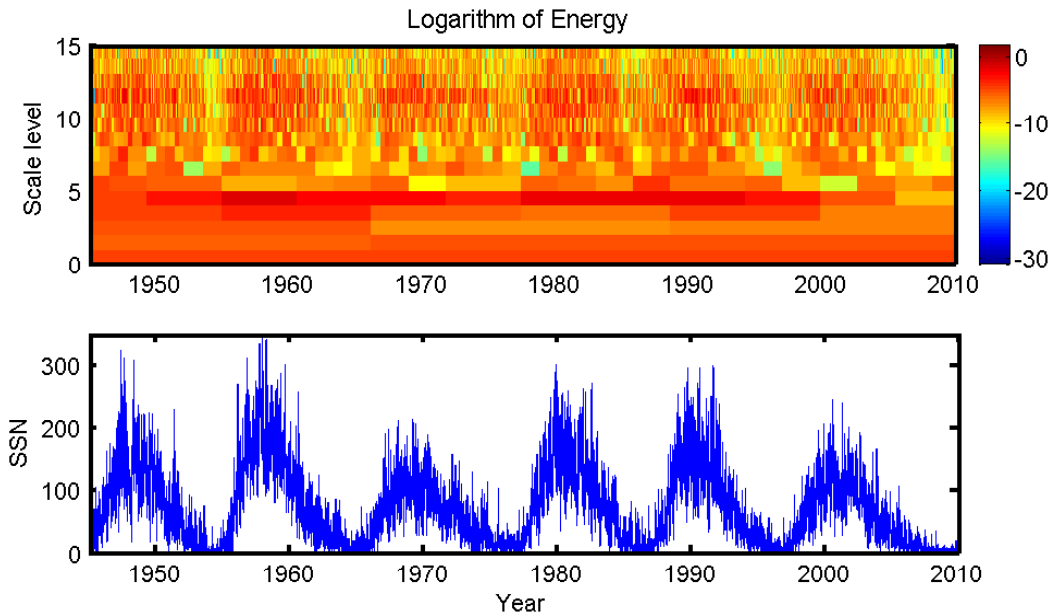}
	\caption{Daily series and orthonormal wavelet spectrum of the 
		sunspot numbers based on the spline wavelets built in
		\cite{LA2007}. 
		\protect{\label{fig:ssn_spectrum}}}
	\end{center}
	\end{figure}

As a final remark, it is interesting to see that the maximum 
information flow from the SSN lagged by 45 years to the Pacific SST 
agrees very well with frequent occurrences of \ENSO Modoki
during the period of 2000-2010: three of the four \ENSO events 
are of this type (i.e., the 02/03, 04/05, 09/10 episodes). 
If we go back to 45 years ago, the sun 
is most active during the period of 1955-1965 (Fig.~\ref{fig:ssn_spectrum}) 
Particularly, in the wavelet spectrum, 
there is a distinct high at the scale level of 4 (corresponding to a time
scale of $2^{-4-(-15)} = 2048$~days~$\approx5.6$~years)---Recall that this
is roughly the sampling interval we used in choosing the time-delay series 
to form the $\dims$-dimensional system and compute the information flow.
After that period, the sun becomes quiet, correspondingly 
we do not see much \ENSO Modoki for the decade 2010-20. But 
two peaks of SSN are seen in the twenty years since 1975-78. 
Does this correspondence
herald that, after entering 2020, in the following two decades,
\ENSO Modoki may frequent the equatorial Pacific again? 
We don't know whether this indeed points to a link, 
but when this project was initiated early in 2018, the projected El
Ni\~no Modoki and La Ni\~na Modoki already occurred recently. 
This question, among others, are to be addressed in future studies.
 (For reference, the codes used in this study are available at\\
   http://www.ncoads.org/enso\_modoki\_data\_codes.tar.gz.
  )

\begin{appendix}
\section*{\MakeUppercase{Method}}

\noindent
{\bf Estimation of information flow and causality among multivariate time series.}
%
Information flow is a fundamental notion in general physics 
which has applications in a wide
variety of scientific disciplines (see \cite{Liang2016} and references therein). 
Its importance lies beyond the literal meaning in that it 
implies causation, uncertainty propagation, predictability transfer, etc.
Though with a history of research for more than 30 years,
it has just been put on a rigorous
footing and derived from first principles, 
initially motivated by the predictability study in atmosphere-ocean
science\cite{LK2005}. 
Since its birth it has been validated with many benchmark dynamical systems
such as baker transformation, H\'enon map, R\"ossler system, etc. 
(cf.~\cite{Liang2016}\cite{Liang2014}), and has been applied with success to different problems 
in earth system science, neuroscience and quantitative finance, e.g., 
\cite{Vannitsem2019}, \cite{Dionisis2019}, \cite{Stips2016}, to name a few.
Hereafter we just give a very brief introduction.


Consider a dynamical system
	\begin{eqnarray}	\label{eq:stoch_gov}
	\dt{\ve x} = \ve F(t; \ve x) + \vve B(t; \ve x) \dot{\ve w},
	\end{eqnarray}
where $\ve x$ and $\ve F$ are $\dims$-dimensional vector, $\vve B$ is a
$\dims$-by-$m$ matrix, and $\dot{\ve w}$ an $m$-vector of white noise.
Here we follow the convention in physics and do not distinguish 
the notation of a random variable from that of a deterministic variable. 
Liang (2016)\cite{Liang2016} 
established that the rate of uncertainty (in terms of Shannon entropy)
transferred from $x_2$ to $x_1$, or rate of information flowing from 
$x_2$ to $x_1$ (in nats per unit time), is: 
%
%
	\begin{eqnarray}	\label{eq:Tn21}
	T_{2\to1} 
	= -E \bracket{\frac1{\rho_1} 
	       \int_{\R^{\dims-2}} \DI{(F_1\rho_{\excl2})} dx_3...dx_\dims}
	    + \frac12 E \bracket{\frac1{\rho_1} 
 	   \int_{R^{\dims-2}} \DIDI {(g_{11}\rho_{\excl2})} dx_3...dx_\dims},
	\end{eqnarray}
	where 
	 $E$ stands for mathematical expectation, 
	 $\rho_1 = \rho_1(x_1)$ is the marginal probability density function
	 (pdf) of $x_1$, 
	$\rho_{\excl2} = \int_\R \rho dx_2$, 
	and $g_{11} = \sum_{j=1}^m b_{1j} b_{1j}$.
The units are in nats per unit time.
Ideally if $T_{2\to1} = 0$, then $x_2$ is not causal to $x_1$; otherwise it
is causal (for either positive or negative information flow). But in
practice significance test is needed. 

Generally $T_{2\to1}$ depends on $(x_3,...,x_\dims)$ as well as $(x_1,
x_2)$. But it has been established that\cite{Liang2018}
	{\it
	it is invariant upon arbitrary 
	nonlinear transformation of $(x_3,x_4,...,x_\dims)$,
	}
indicating that information flow is an intrinsic physical property.
Also established is the {\it principle of nil causality}, which asserts
that, if the evolution of $x_1$ is independent of $x_2$, then
$T_{2\to1}=0$. This is a quantitative fact that all causality analyses try
to verify in applications, while in the framework of information flow
it is a proven theorem.

In the case of linear systems where $\ve F(\ve x,t) = \vve A \ve x$, 
$\vve A = (a_{ij})$, the formula (\ref{eq:Tn21}) becomes quite simple\cite{Liang2016}:
	\begin{eqnarray}	\label{eq:T21_linear}
	T_{2\to1} = a_{12} \frac {\sigma_{12}} {\sigma_{11}},
	\end{eqnarray}
where $\sigma_{ij}$ is the population covariance.
An immediate corollary is that {\it causation implies correlation, while
correlation does not imply causation},
fixing the long-standing philosophical debate over causation versus 
correlation ever since Berkeley (1710)\cite{Berkeley1710}. 

To arrive at a practically applicable formula, we need to estimate
(\ref{eq:T21_linear}), given $\dims$ time series. The estimation roughly
follows that of Liang (2014)\cite{Liang2014}, and can be found in
\cite{Liang2021}. The following is just a brief summary of some relevant
pieces as detailed in \cite{Liang2021}.

Suppose we have $\dims$ time series, $x_i$, $i=1,...,\dims$, and these
series are equi-spaced, 
all having $N$ data points $x_i(n)$, $n=1,2,...,N$.
Assume a linear model 
$\ve F(\ve x;t) = \vve A \ve x$, $\vve A = (a_{ij})$ being
a $\dims\times\dims$ matrix), 
and $\vve B$ being a $\dims\times\dims$ diagonal matrix.
To estimate $T_{2\to1}$, we first need to estimate $a_{12}$.
As shown before in \cite{Liang2014}, 
when $a_{ij}$ and $b_i$ are constant, the maximal
likelihood estimator (mle) is precisely the least square solution of
the following $N$ (overdetermined) algebraic equations
	\begin{eqnarray}
	\sum_{j=1}^\dims a_{1j} x_j(n) = \dot x_1(n),\qquad\qquad n=1,...,N
	\end{eqnarray}
where $\dot x_1(n) = (x_1(n+1) - x_1(n)) / \Delta t$ is the differencing
approximation of $dx_1/dt$ using the Euler forward scheme,
and $\Delta t$ is
the time stepsize (not essential; only affect the units).
Following the procedure in \cite{Liang2014}, 
the least square solution 
of $(a_{11},..., a_{1\dims})$, 
$(\hat a_{11}, ..., \hat a_{1\dims})$, 
satisfies the algebraic equation
	\begin{eqnarray*}
	\matthree {C_{1,1}}  {...}  {C_{1,\dims}}
		    \vdots		 \vdots		\vdots
	          {C_{\dims,1}}  {...}  {C_{\dims,\dims}}
	\vecthree {\hat a_{11}}  {\vdots}  {\hat a_{1\dims}}
	=
	\vecthree {C_{1,d1}}
			\vdots
	          {C_{\dims,d1}},
	\end{eqnarray*}
where 
	\begin{eqnarray*}
	&&C_{ij} = \frac1N \sum_{n=1}^N 
		   (x_i(n) - \bar x_i) (x_j(n) - \bar x_j) \\
	&&C_{i,dj} = \frac1N \sum_{n=1}^N (x_i(n) - \bar x_i) 
			       (\dot x_j(n) - {\bar {\dot x}}_j)
	\end{eqnarray*}
are the sample covariances.
Hence
	$\hat a_{12} = \frac 1 {\det\vve C} \cdot 
		       \sum_{j=1}^\dims \Delta_{2j} C_{j,d1}$
where $\Delta_{ij}$ are the cofactors.
This yields an estimator of the information flow from $x_2$ to $x_1$:
	\begin{eqnarray}	 \label{eq:T21_hat}
	\hat T_{2\to1} = \frac 1 {\det\vve C} \cdot 
		       \sum_{j=1}^\dims \Delta_{2j} C_{j,d1}
			\cdot \frac {C_{12}} {C_{11}},
	\end{eqnarray}
i.e., Eq.~(\ref{eq:T21_hat}) in section~2.
Here $\det\vve C$ is the determinant of the covariance matrix $\vve C$, and
now $T$ is understood as the MLE of $T$. We slightly abuse notation for
the sake of simplicity.)
For two-dimensional (2D) systems, $\dims=2$, the equation reduces to 
	\begin{eqnarray}	\label{eq:T21_2d}
	\hat T_{2\to1} = \frac {C_{11}C_{12} C_{2,d1} - C_{12}^2 C_{1,d1}} 
			  {C_{11}^2 C_{22} - C_{11}C_{12}^2},
	\end{eqnarray}
recovering the familiar one as obtained in \cite{Liang2014} and frequently
used in applications (e.g., \cite{Stips2016}, \cite{Vannitsem2019}, 
\cite{Dionisis2019}).

{
The significance of $\hat T_{2\to1}$ of (\ref{eq:T21_hat}) can be tested
following the same strategy as that used in \cite{Liang2014}. 
By the MLE property (cf.~\cite{Garthwaite1995}, when $N$ is large, 
$\hat T_{2\to1}$ approximately follows a Gaussian around its true 
value with a variance $\parenth{\frac{C_{12}}{C_{11}}}^2
{\hat\sigma}_{a_{12}}^2$. Here ${\hat\sigma}_{a_{12}}^2$ is 
the variance of $\hat a_{12}$, which is estimated as follows.
   Denote by $\veg\theta$ the vector of parameters to be estimated; here
it is $(a_{11}, a_{12}, a_{13},...,a_{1\dims}; b_1)$. 
Compute the Fisher information matrix $\vve I = (I_{ij})$ 
	\begin{eqnarray*}
	I_{ij} = - \frac 1N \sum_{n=1}^N 
	   \frac {\D^2 \log \rho(\ve x_{n+1} | \ve x_n; \hat{\veg\theta} ) } 
			 {\D\theta_i \D\theta_j}
	\end{eqnarray*}
where $\rho(\ve x_{n+1} | \ve x_n)$ is a Gaussian for a linear model:
	\begin{eqnarray*}
	\rho(\ve x_{n+1} | \ve x_n)
	&&= \frac 1 {[(2\pi)^2\det (\vve B \vve B^T \Delta t )]^{1/2}} \cr
	&&\qquad  \times  
	  e^{-\frac12 (\ve x_{n+1}-\ve x_n -\vve A\ve x_n\Delta t)^T  
	              (\vve B\vve B^T\Delta t)^{-1}
	              (\ve x_{n+1}-\ve x_n -\vve A\ve x_n\Delta t)
	    }.
	\end{eqnarray*}
The computed entries are then evaluated using the estimated parameters, and
hence the Fisher information matrix is obtained. It has been established
(cf.~\cite{Garthwaite1995}) that
	$\vve I^{-1}/N$
is the covariance matrix of $\hat{\veg\theta}$, from which it is easy to
find the entry $\hat\sigma_{a_{12}}^2$ [here it is the entry (2,2)]. 
Given a level, the significance 
of an estimated information flow can be tested henceforth.
}

\subsection*{Data availability}
The data used in this study include the SST and sunspot number (SSN) from 
National Oceanic and Atmospheric Administration (NOAA)
(SST: http://www.esrl.noaa.gov/psd/data/gridded/cobe2.html;
SSN:
https://wwww.esrl.noaa.gov/psd/gcos\_wgsp/Timeseries/Data/sunspot.long.data),
and the \ENSO Modoki Index (EMI) from Japan Agency for Marine-Earth Science
and Technology (JAMSTEC)
(http://www.jamstec.go.jp/aplinfo/sintexf/e/elnmodoki/data.html).
Among them SST has a horizontal resolution of $1^{\rm o}\times1^{\rm o}$.
Temporally all of them are monthly data, and
at the time when this study was initiated, SST, EMI, and SSN respectively
have a time coverage of 01/1850-12/2018, 01/1870-11/2018, and 05/1847-12/2017.
%
%
Daily SSN is also used for spectral analysis; it covers the period 
15/04/1921-31/12/2010.

\subsection*{Code availability}
All the codes (and some of the downloaded data) are available at\\
http://www.ncoads.org/upload/enso\_modoki\_data\_codes.tar.gz \\

\noindent
{\bf Acknowledgments.}
The long discussion with Richard Lindzen on August 17, 2019,
and the email exchange later on in late February 2020 are sincerely
appreciated.
The comments from Dmitry Smirnov and 
four anonymous reviewers have helped improve the manuscript.
We thank NOAA and JAMSTEC for making the data available, and
the NUIST High Performance Computing Center for 
providing computational resources.
This study is partially funded by National Science Foundation of China
(No.~41975064), and the 2015 Jiangsu Program for Innovation Research and
Entrepreneurship Groups. 
\\

\noindent
{\bf Author contributions.}
 XSL: idea, methodology, experiment design, computation (causality analysis
and prediction), writing; 
 FX: experiment design, computation (causality analysis); 
 YR: experiment design, computation (prediction).
 All authors discussed the study results.  \\

\noindent
{\bf Competing interests.}
 The authors declare no competing interests.

\end{appendix}

\newpage

\begin{center}
\section*{
\ENSO Modoki thus far can be mostly predicted more than 10 years ahead of
time}
X. San Liang, \ Fen Xu, \ \& Yineng Rong
\end{center}

\vskip 5cm

\begin{center}
{\Huge\bf
Supplementary Information}
\end{center}

\newpage

\let\origthefigure\thefigure

\linenumbers


\newpage

\renewcommand{\thefigure}{S\origthefigure}
\setcounter{figure}{0}

	\begin{figure}[h]
	\begin{center}
	\includegraphics[angle=0,width=0.45\textwidth, height=0.35\textwidth]
 		{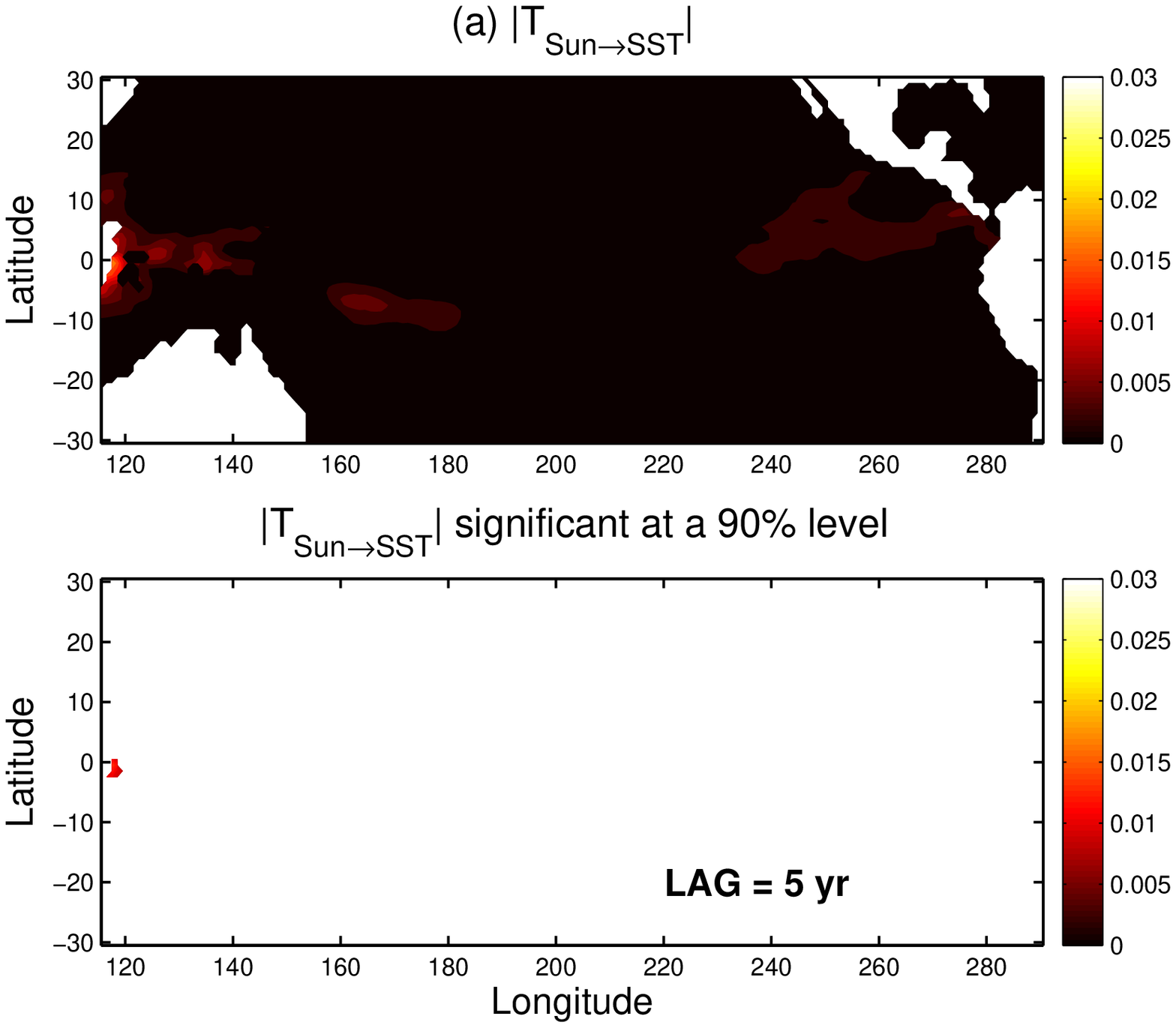}
	\includegraphics[angle=0,width=0.45\textwidth, height=0.35\textwidth]
 		{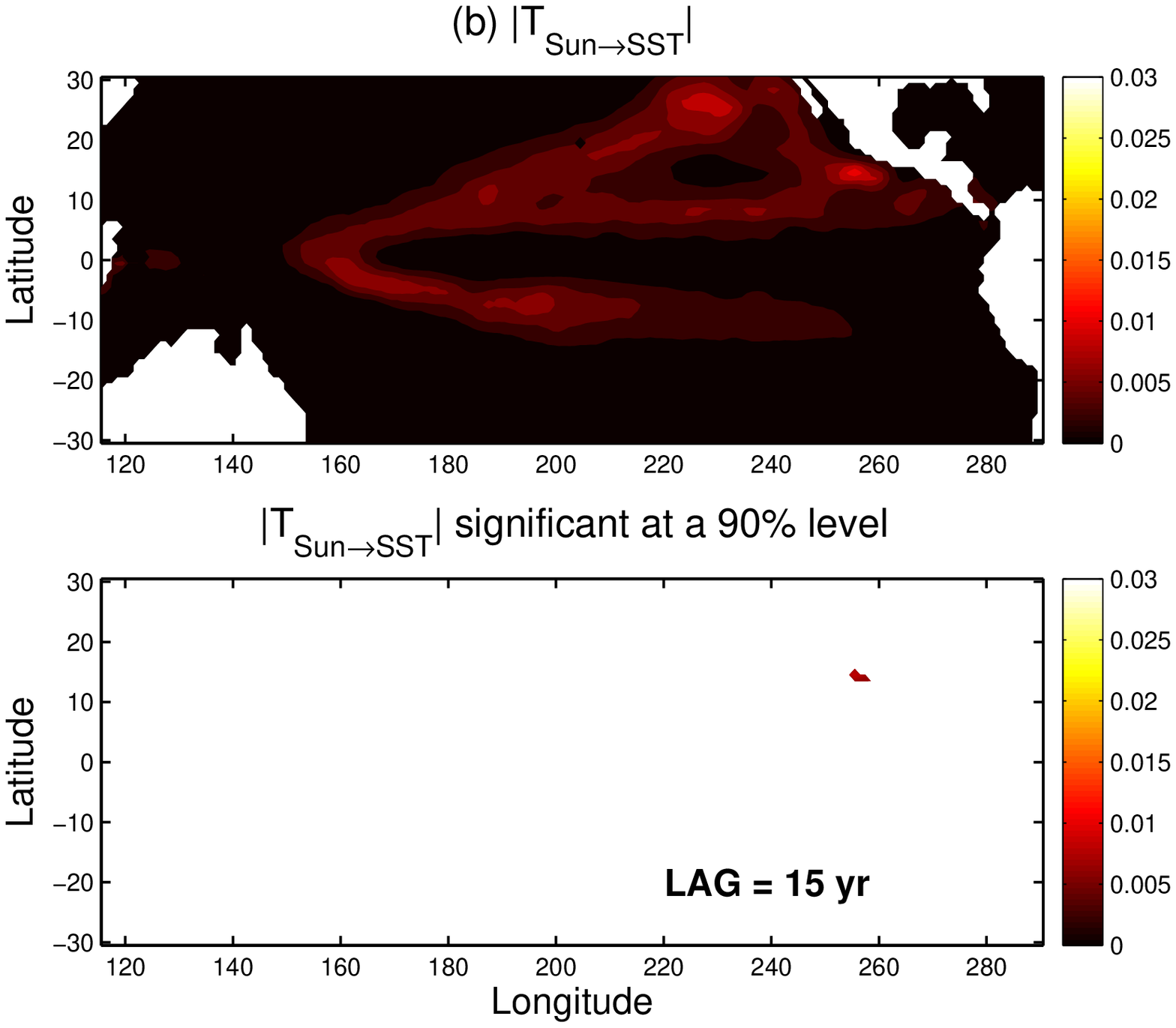}
	\includegraphics[angle=0,width=0.45\textwidth,height=0.35\textwidth]
 		{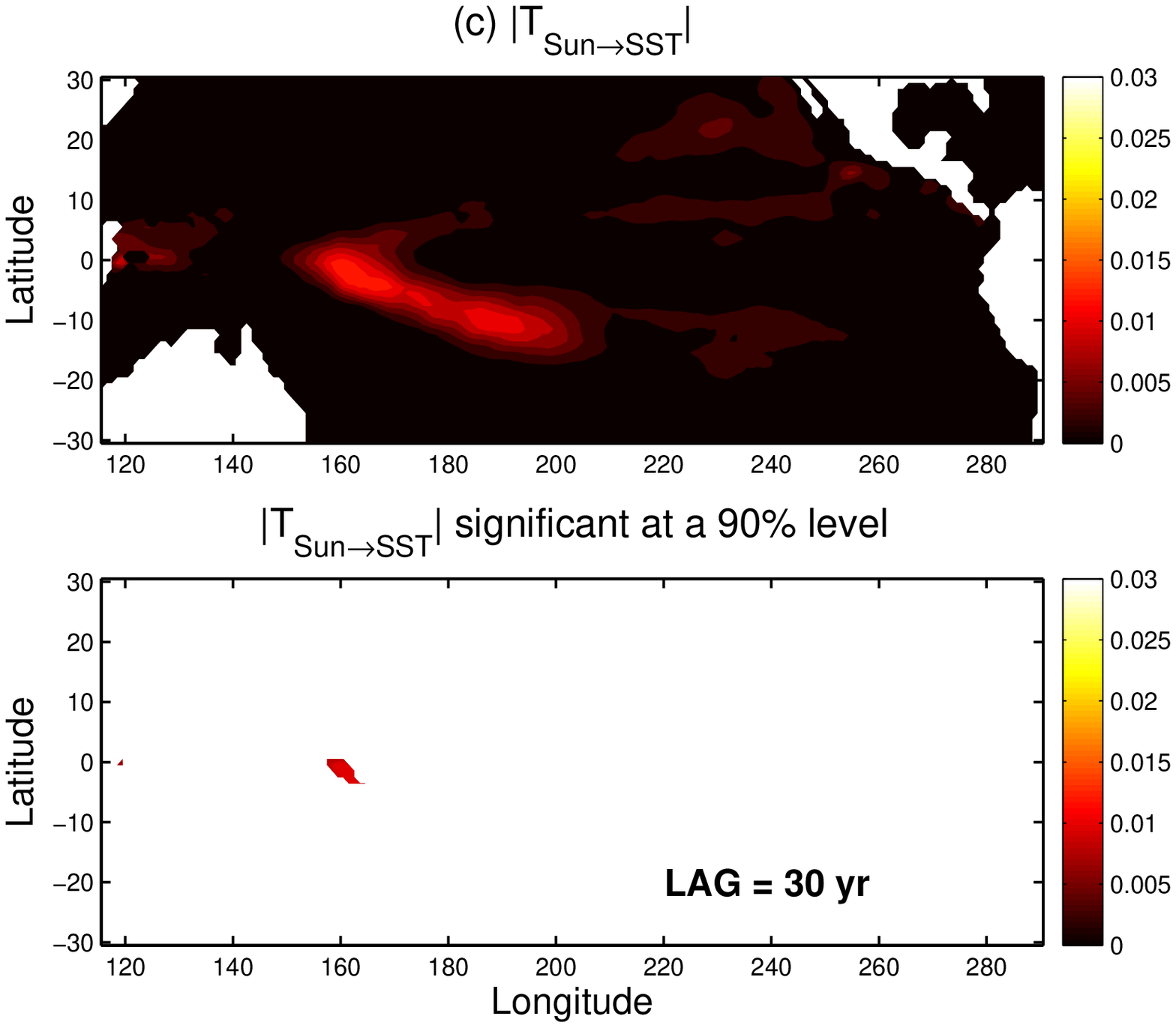}
	\includegraphics[angle=0,width=0.45\textwidth,height=0.35\textwidth]
 		{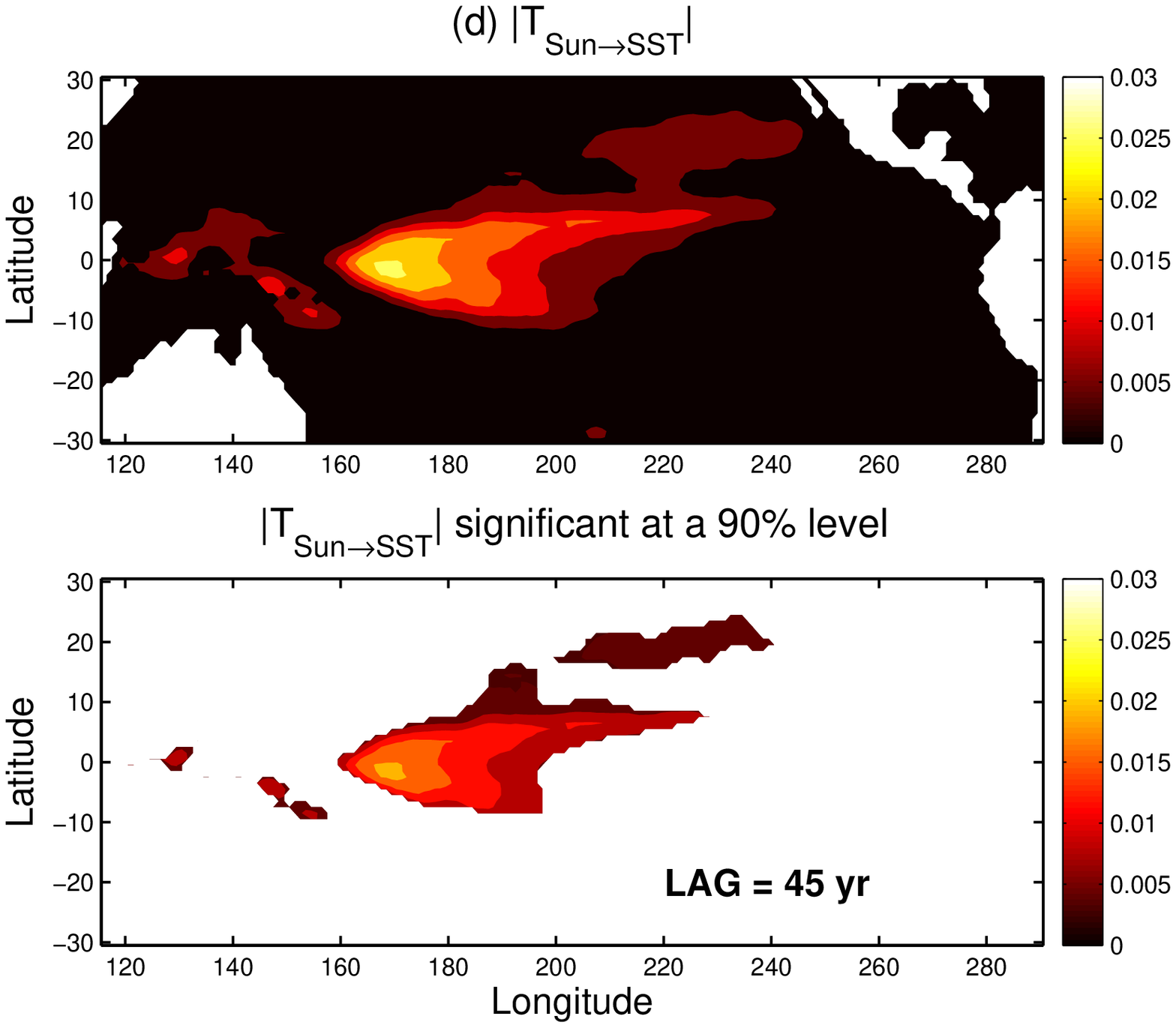}
	\includegraphics[angle=0,width=0.45\textwidth,height=0.35\textwidth]
 		{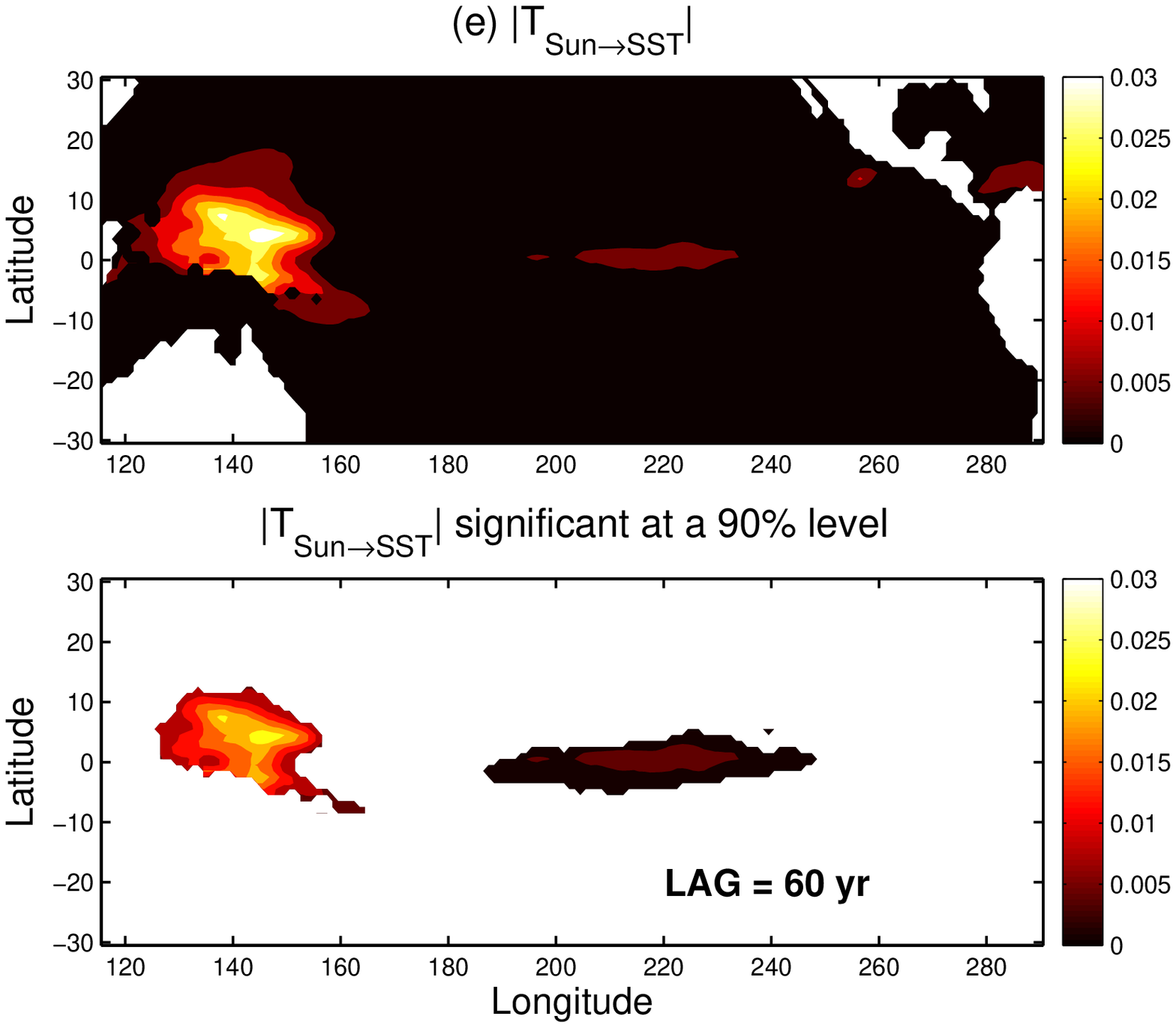}
	\includegraphics[angle=0,width=0.45\textwidth,height=0.35\textwidth]
 		{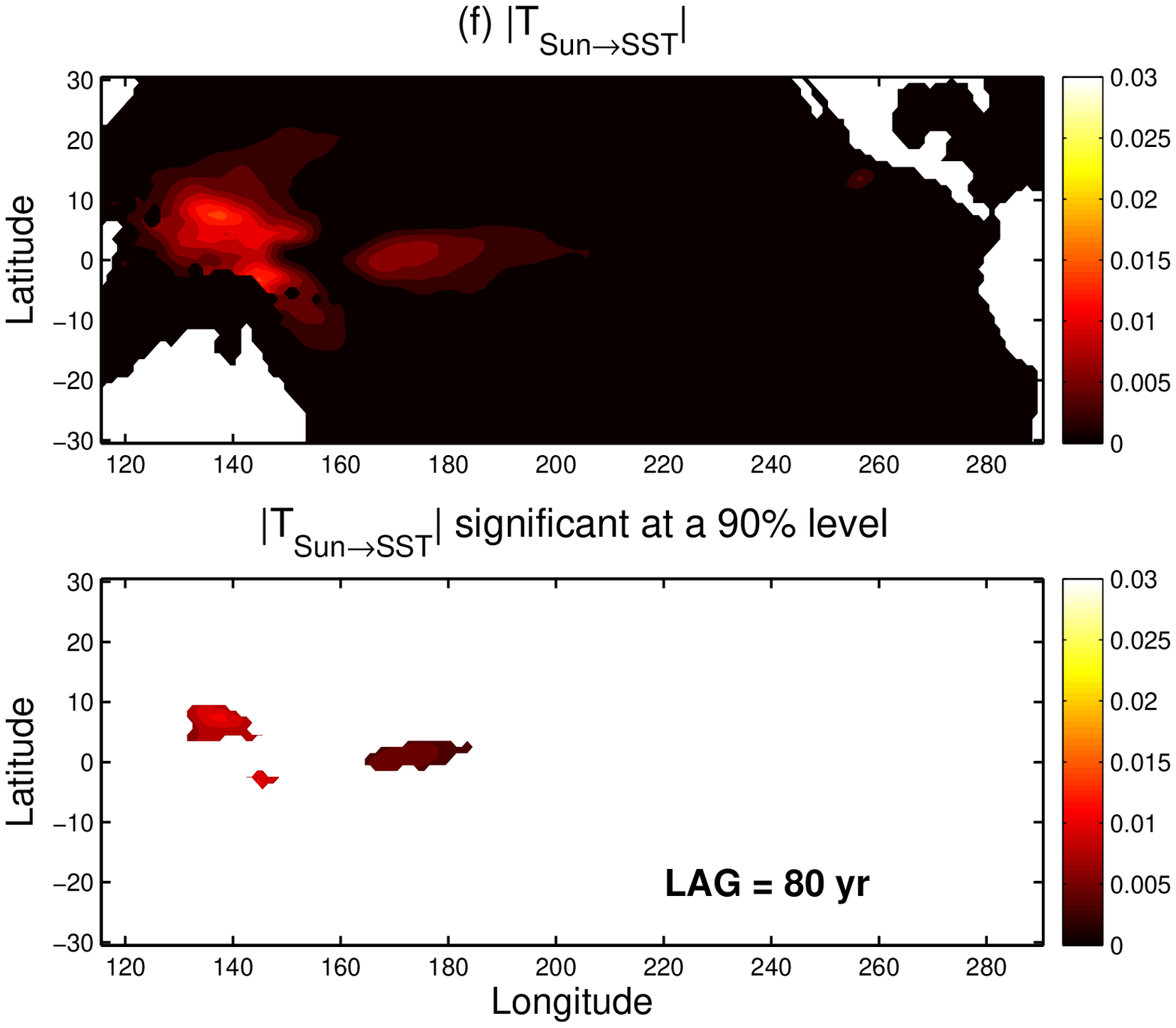}
	\includegraphics[angle=0,width=0.45\textwidth,height=0.35\textwidth]
 		{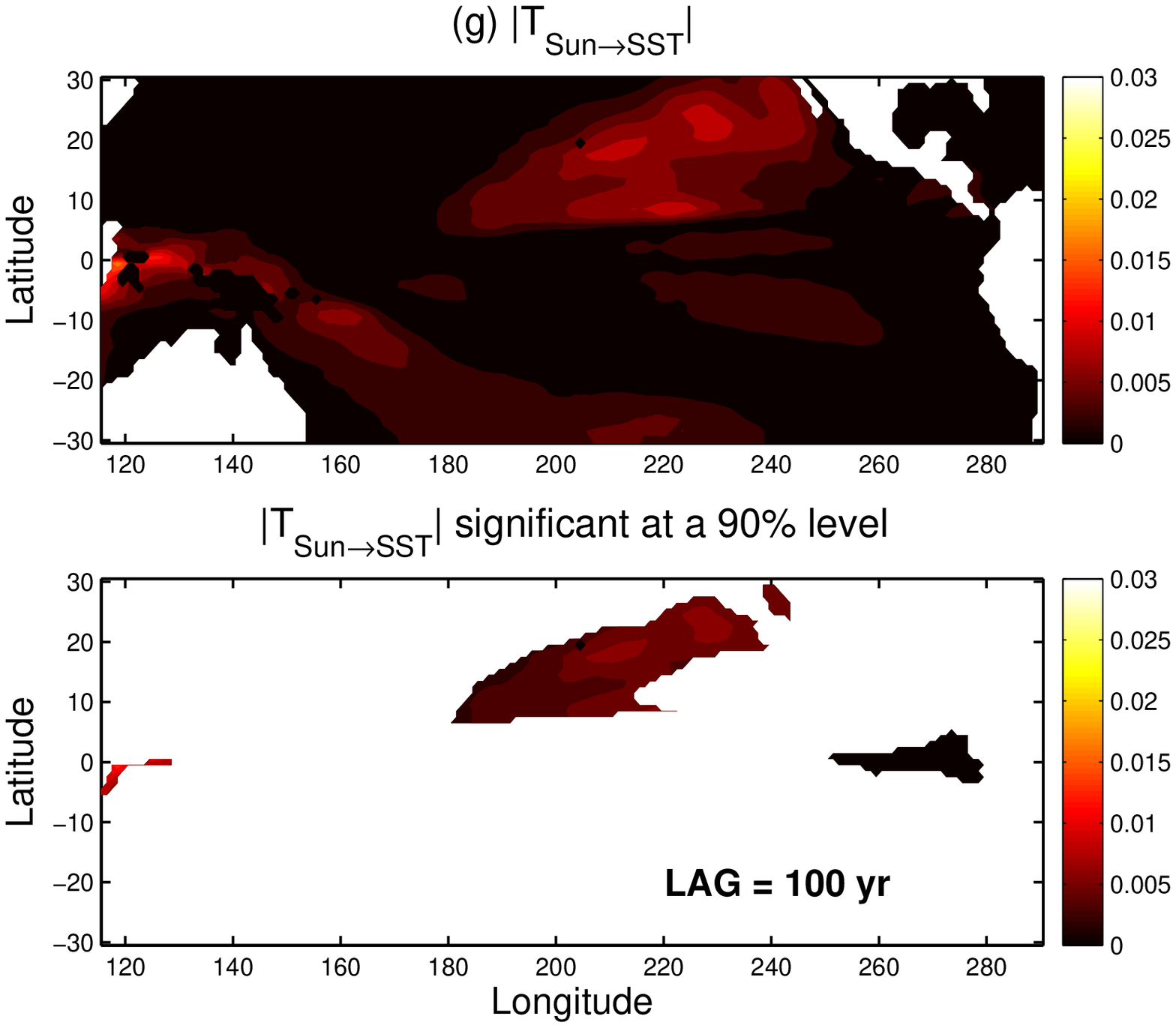}
	\includegraphics[angle=0,width=0.45\textwidth,height=0.35\textwidth]
 		{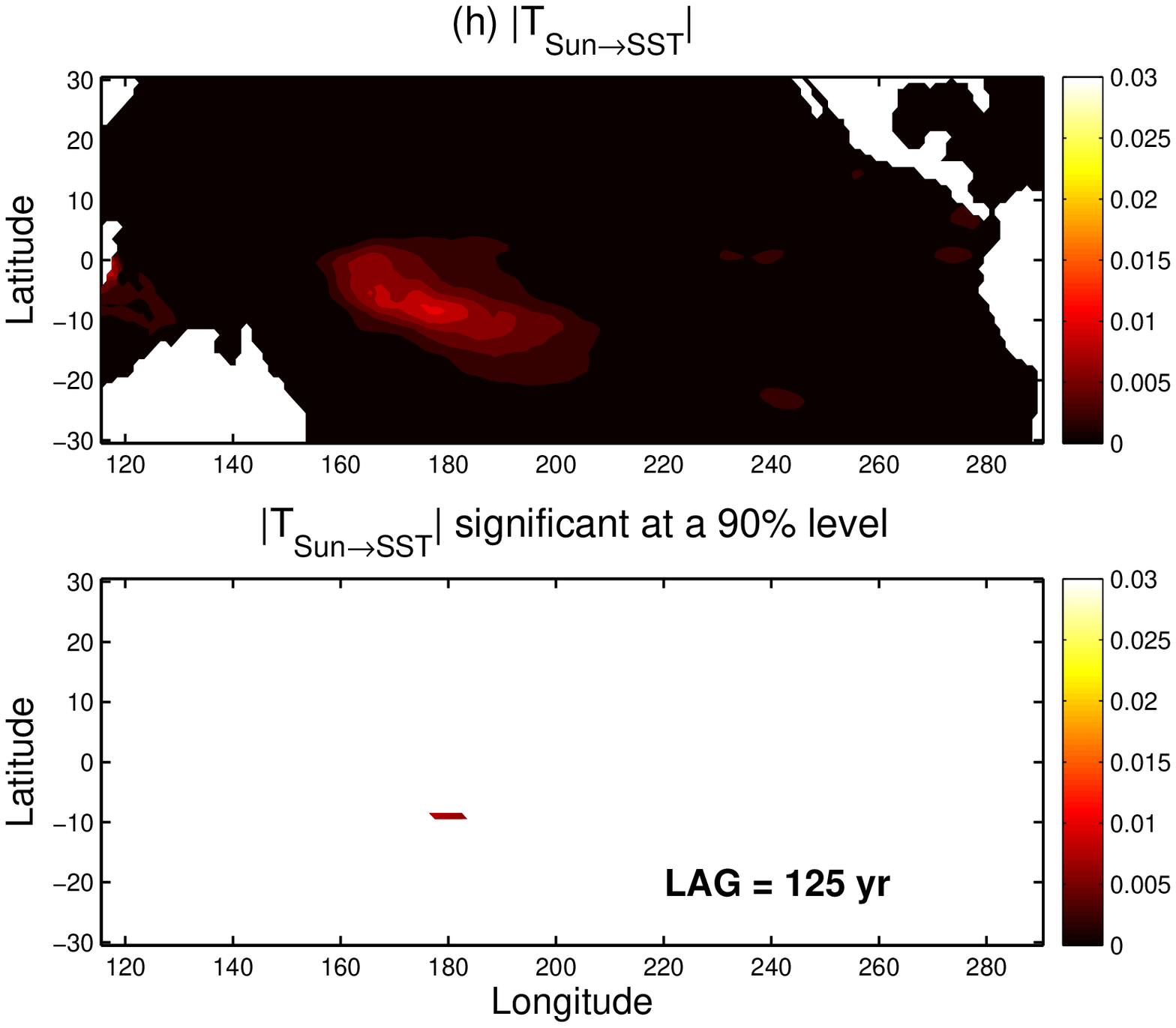}
	\caption{The absolute information flow estimated using the
bivariate formula from the delayed series of sunspot numbers (SSN) to those
of the sea surface temperature (SST) in the Pacific Ocean (in nats/month).
The SST time series between January 1980 - December 2017 are used. In each
subplot, the lower panel shows only the value that is significant at a 90\%
confidence level, and indicated on it is the delay (in years). 
 \protect{\label{fig:S1}}}
	\end{center}
	\end{figure}

	\begin{figure}[h]
	\begin{center}
	\includegraphics[angle=0,width=0.45\textwidth, height=0.35\textwidth]
 		{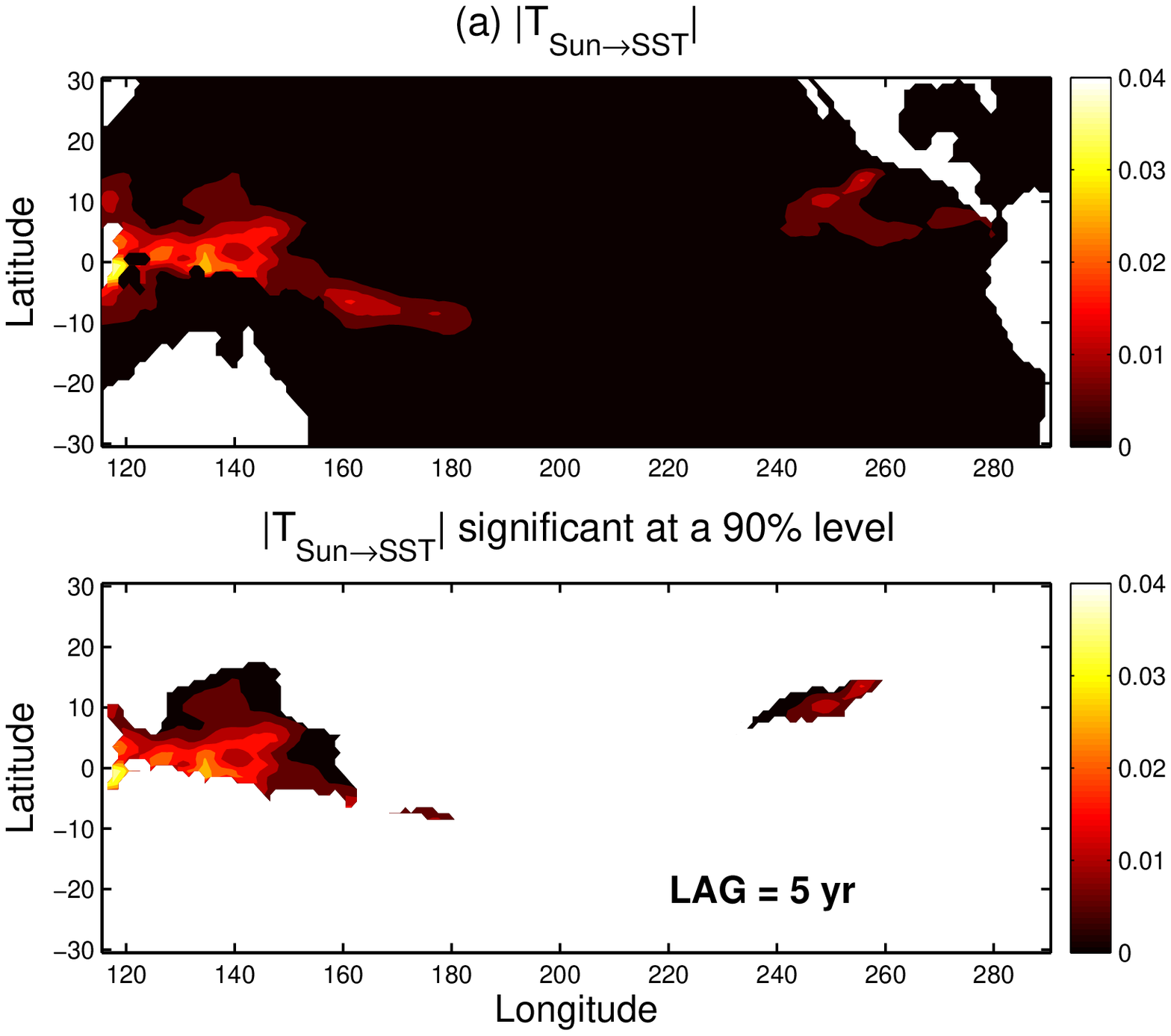}
	\includegraphics[angle=0,width=0.45\textwidth, height=0.35\textwidth]
 		{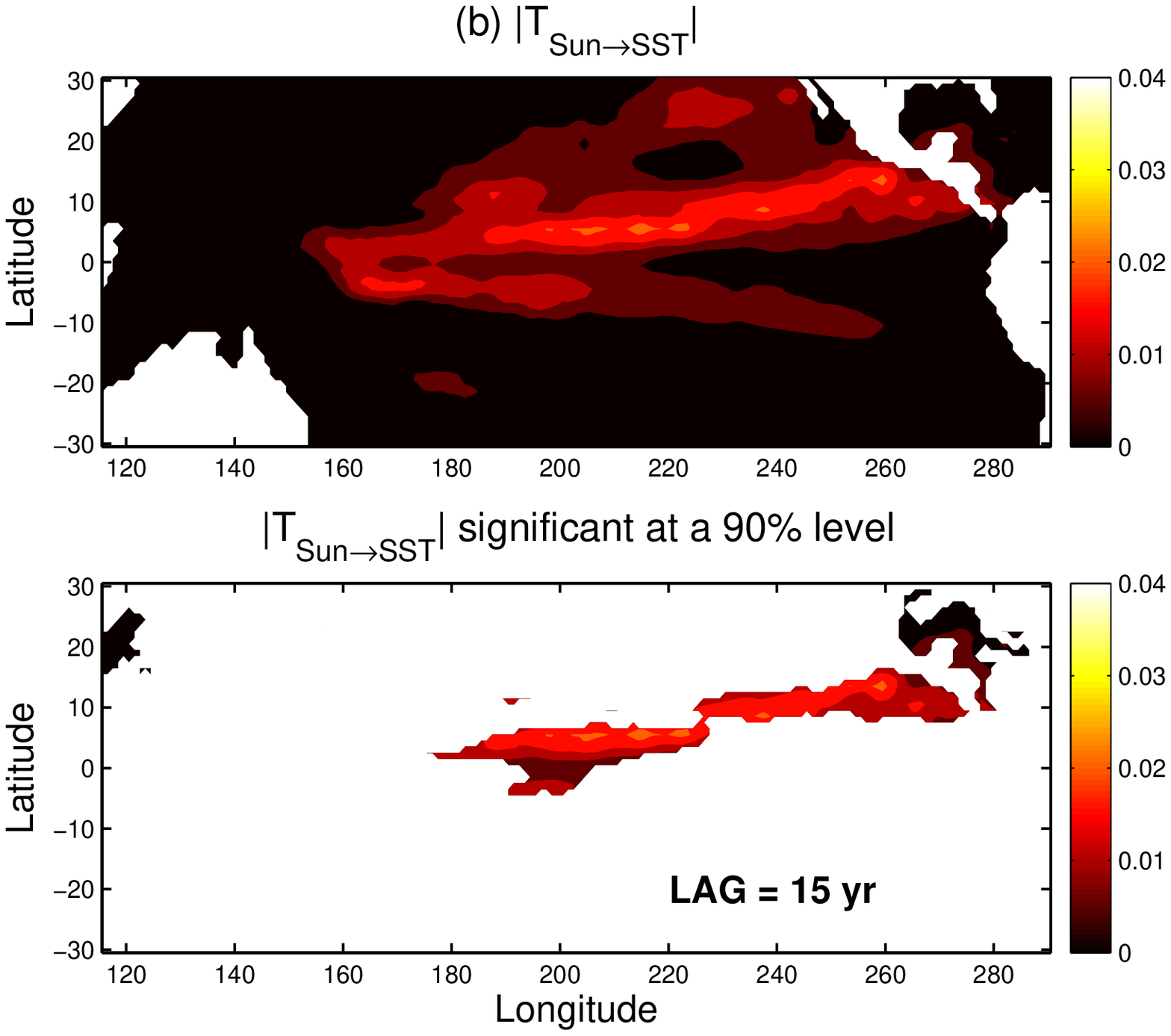}
	\includegraphics[angle=0,width=0.45\textwidth,height=0.35\textwidth]
 		{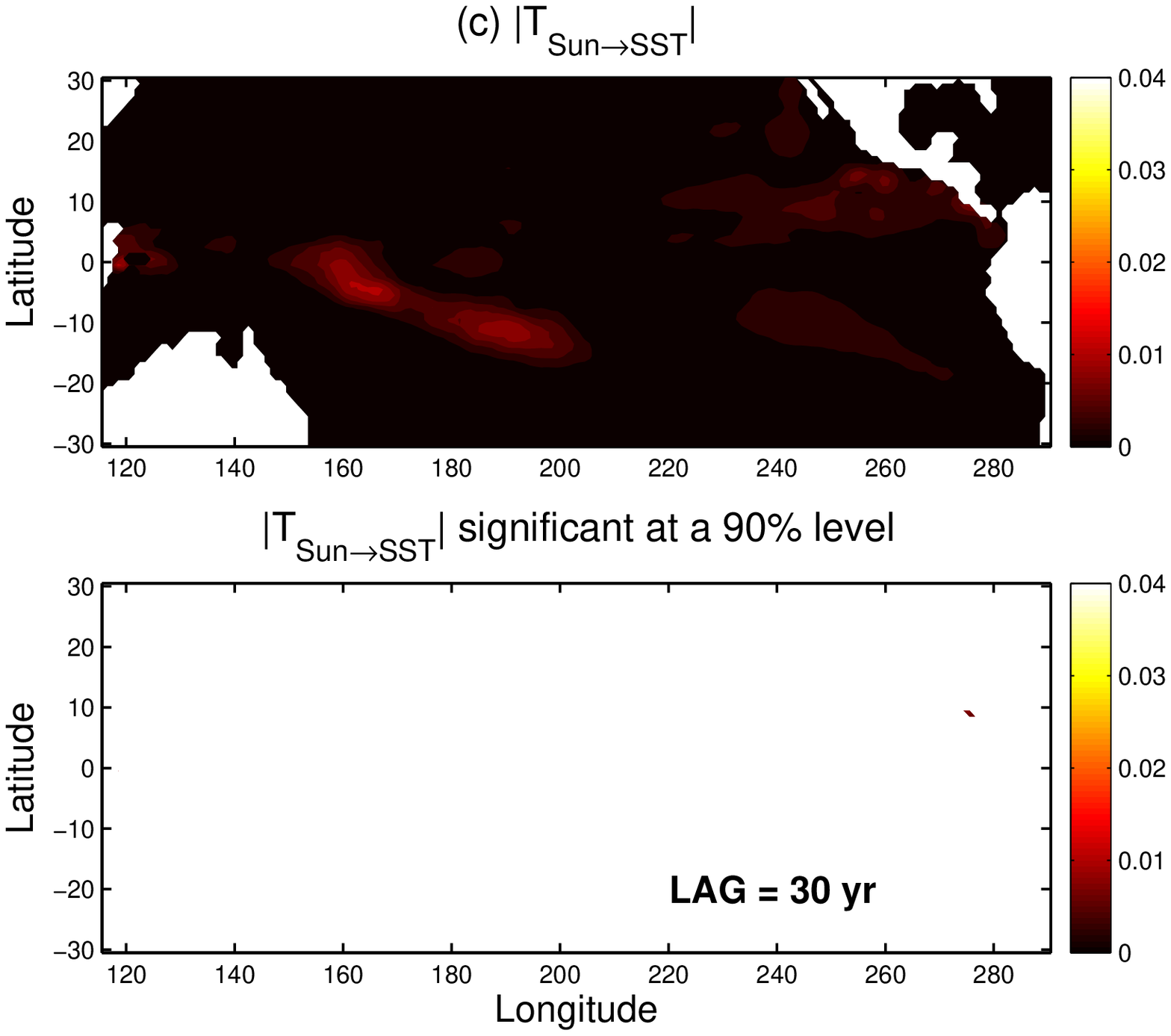}
	\includegraphics[angle=0,width=0.45\textwidth,height=0.35\textwidth]
 		{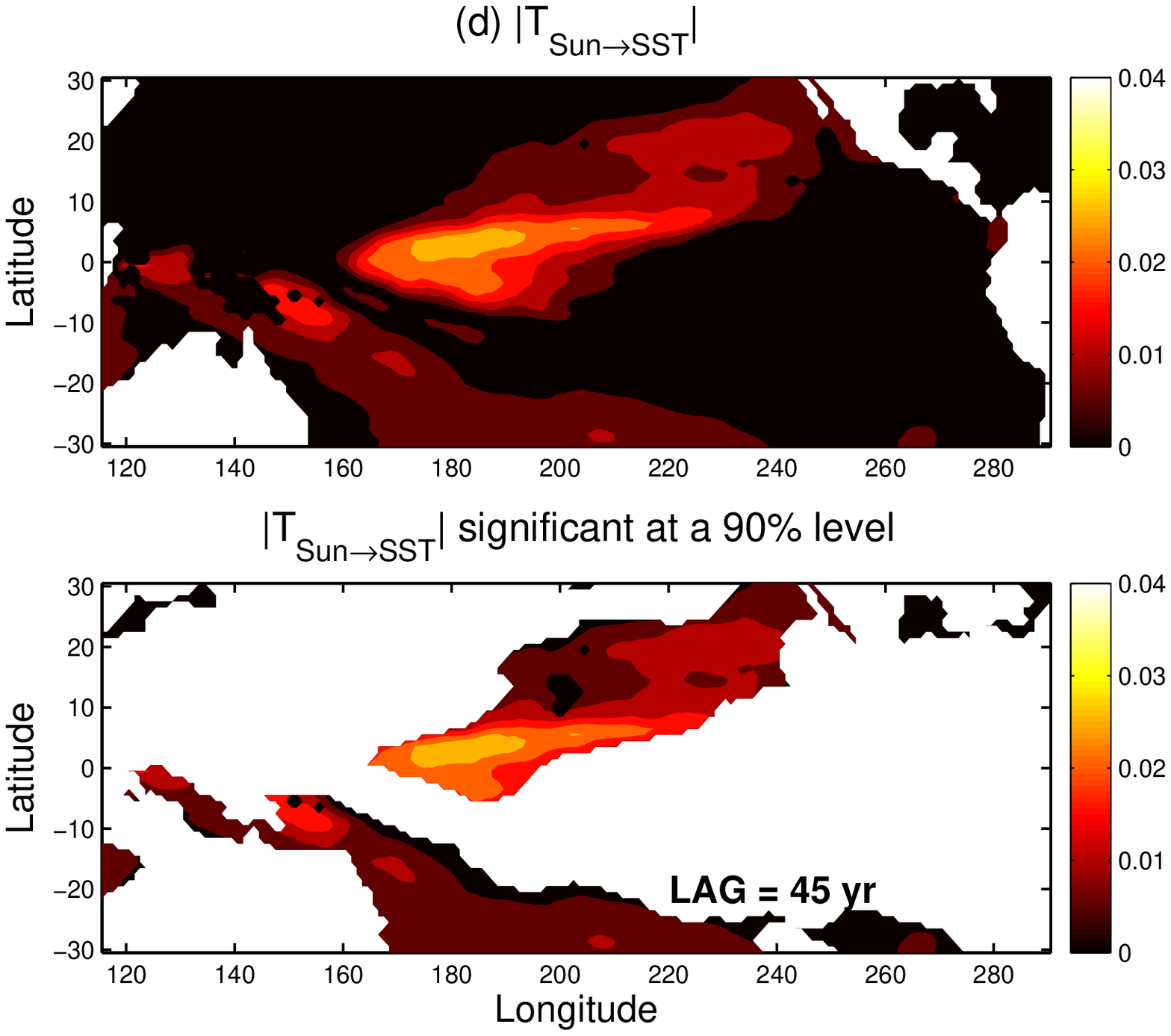}
	\includegraphics[angle=0,width=0.45\textwidth,height=0.35\textwidth]
 		{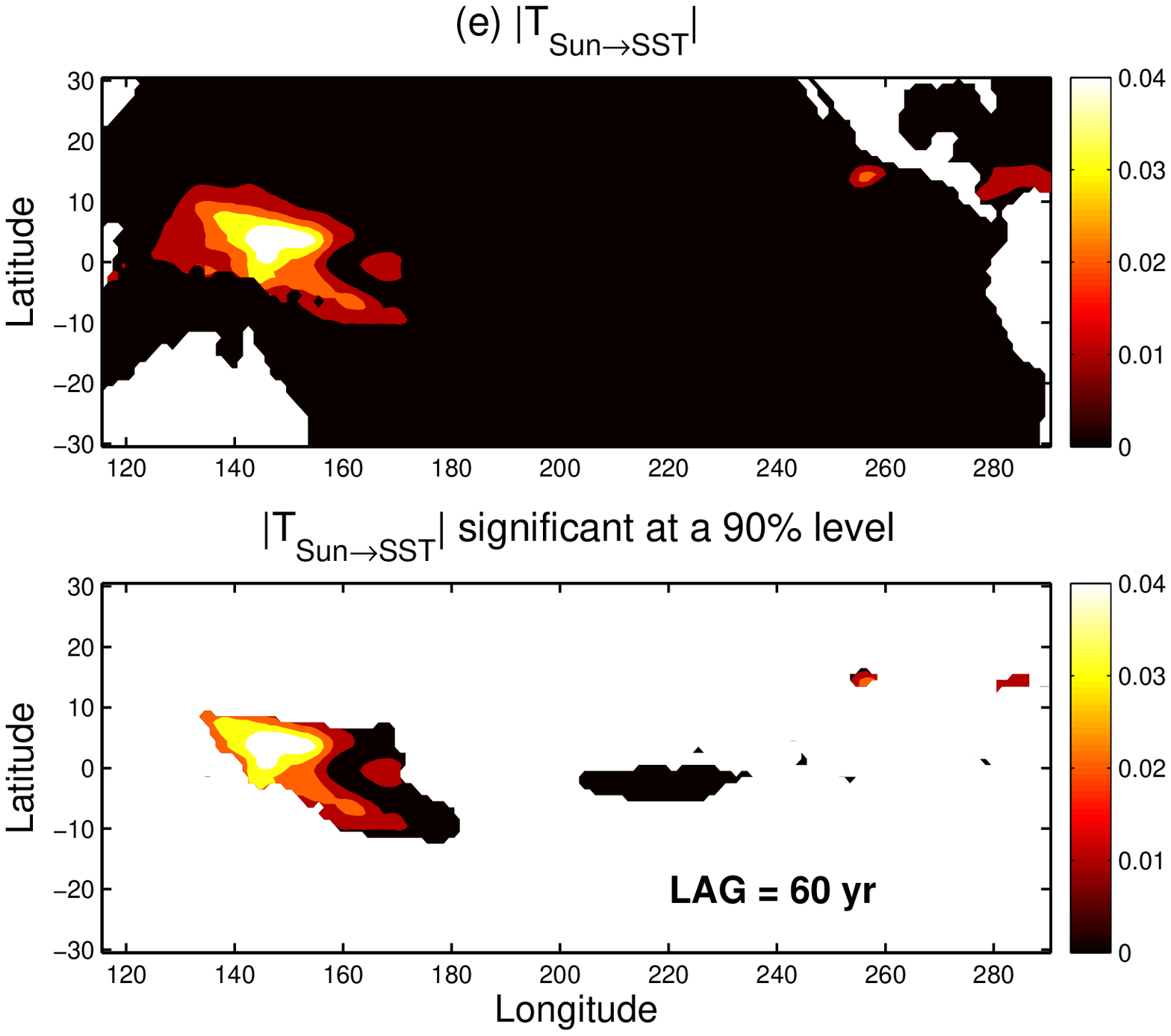}
	\includegraphics[angle=0,width=0.45\textwidth,height=0.35\textwidth]
 		{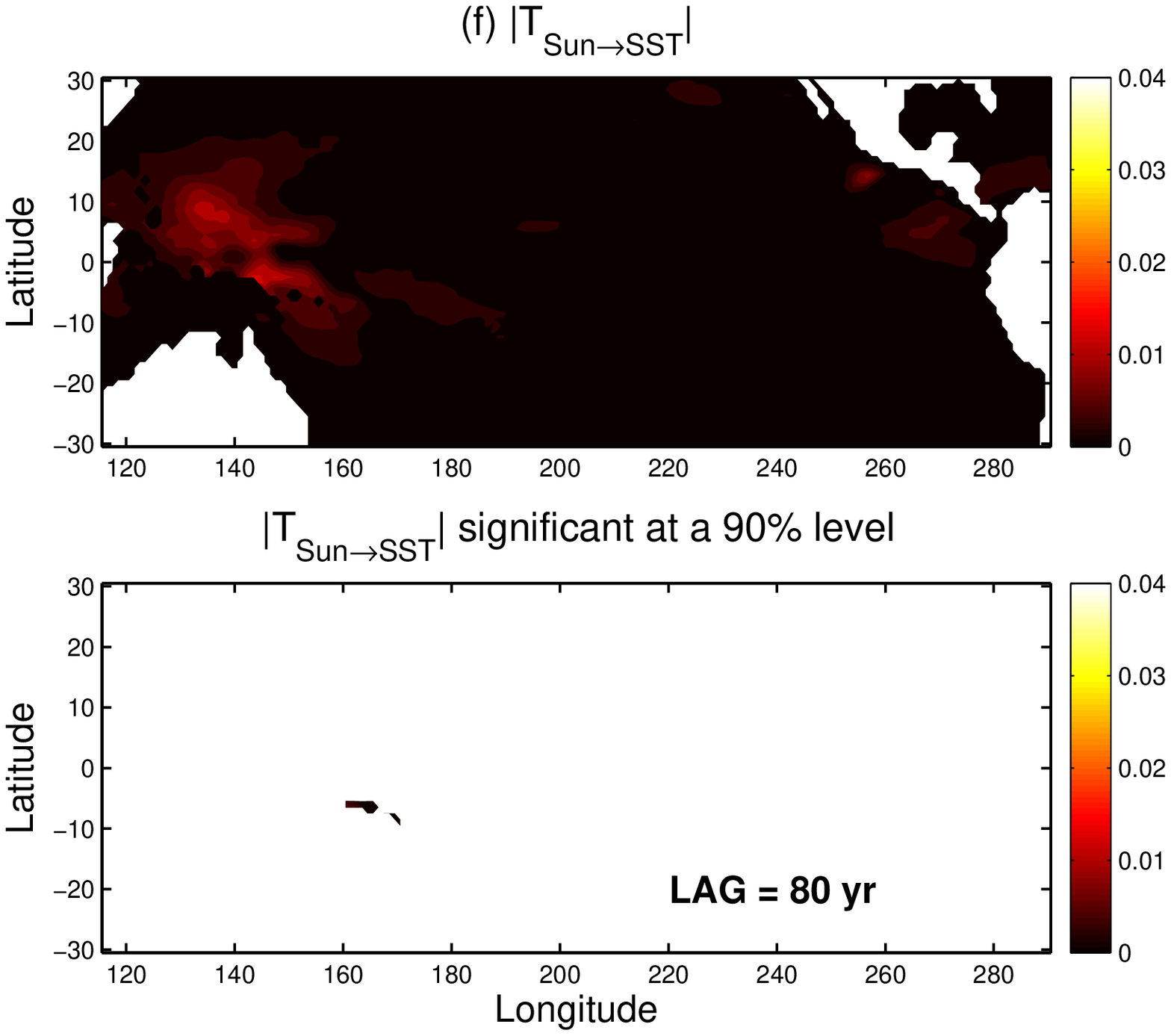}
	\includegraphics[angle=0,width=0.45\textwidth,height=0.35\textwidth]
 		{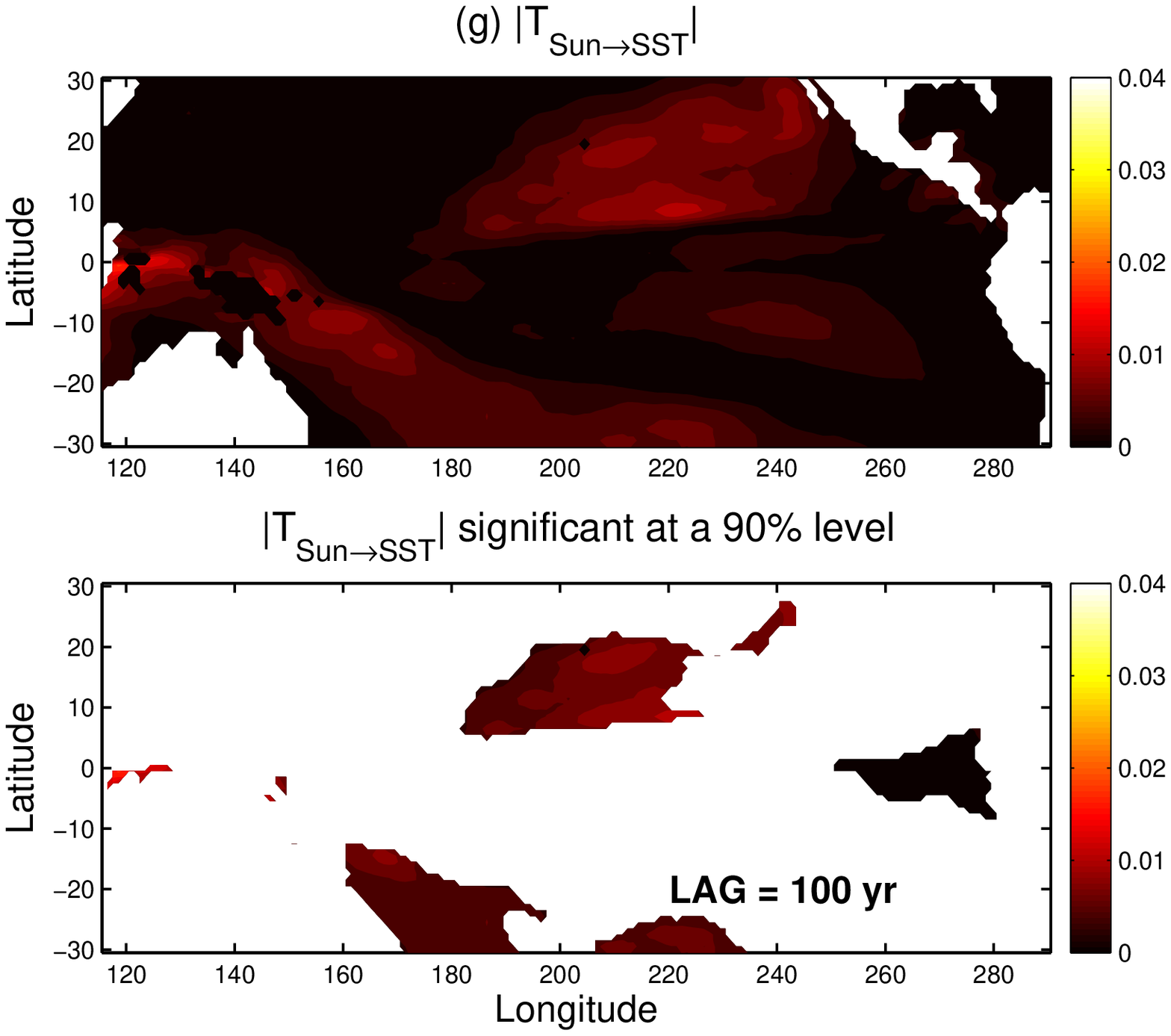}
	\includegraphics[angle=0,width=0.45\textwidth,height=0.35\textwidth]
 		{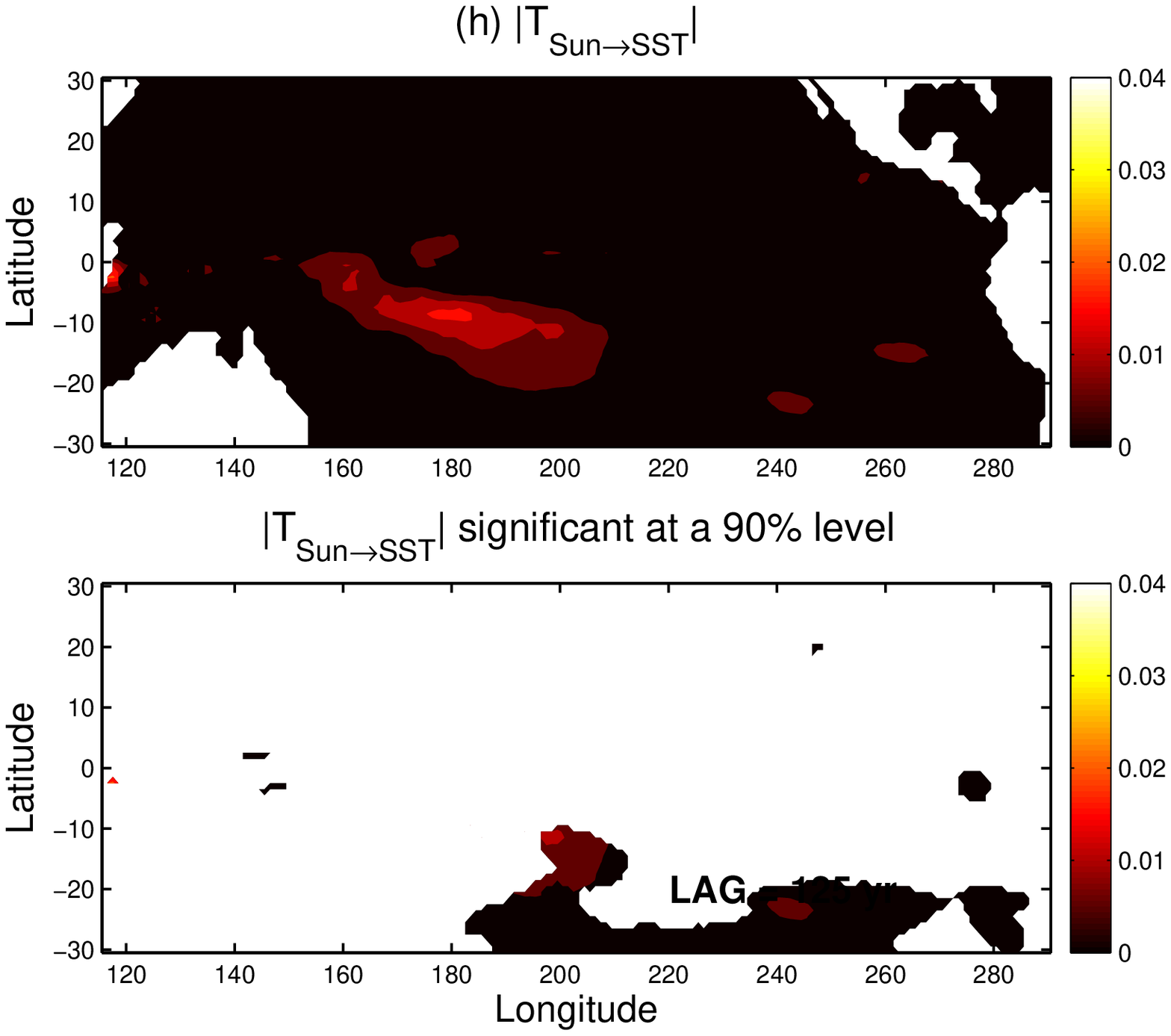}
	\caption{ 
As Supplementary Figure~\ref{fig:S1}, but the information flows are estimated using
the multivariate formula, with the embedding coordinates formed with SSN
series lagged by 22-50 years every 5 years. 
\protect{\label{fig:S2}}}
	\end{center}
	\end{figure}

	\begin{figure}[h]
	\begin{center}
	\includegraphics[angle=0,width=0.8\textwidth]
 		{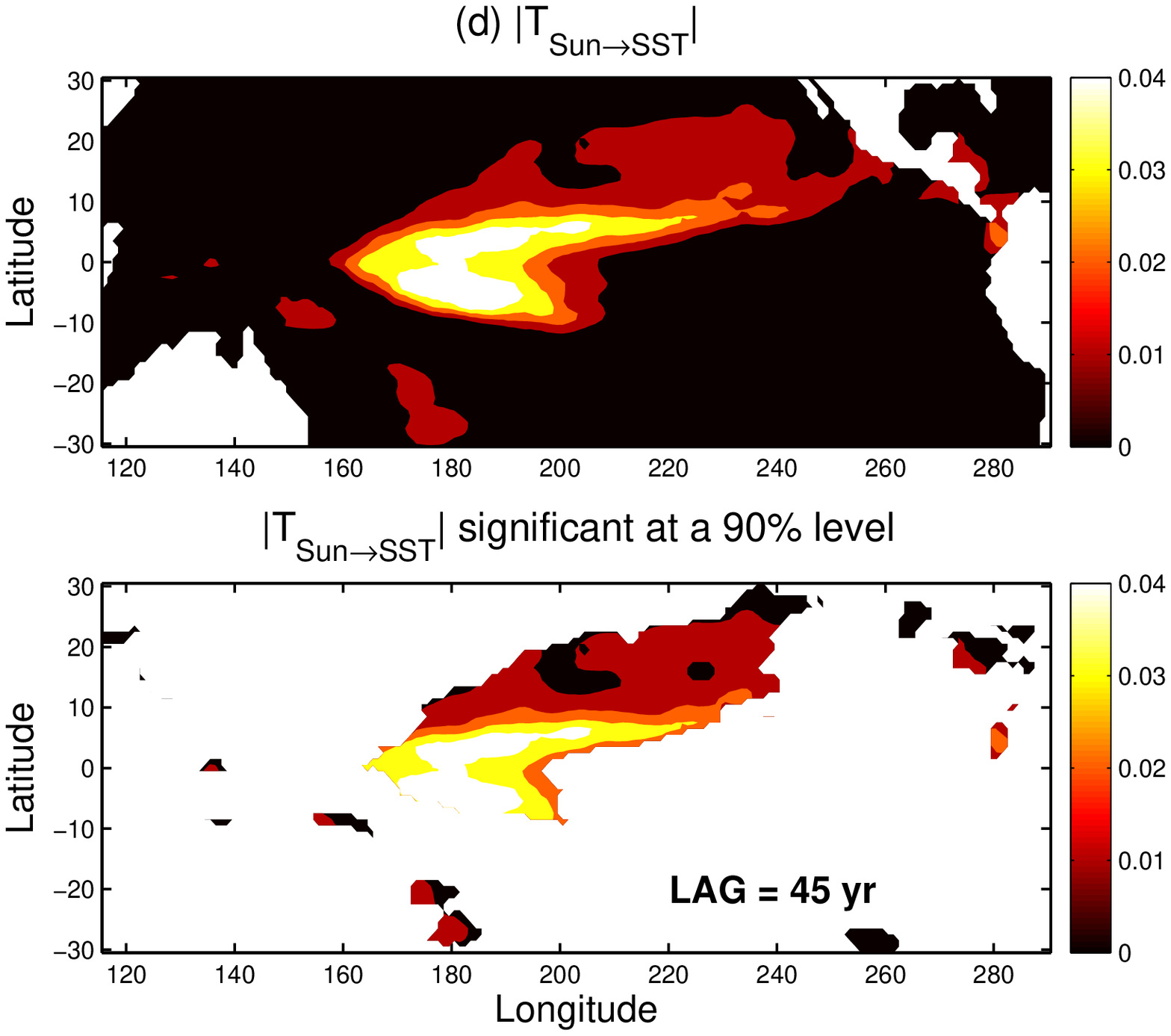}
	\caption{ The absolute information flow from the SSN series delayed
by 45 years to those of the SST in the Pacific Ocean (in nats/month). It is
estimated with the multivariate formula using embedding coordinates formed
with SSN series delayed by 22-50 years every 5 years. The SST time series
between January 1980 - December 2005 are used. In each subplot, the lower
panel shows only the value that is siginificant at a 90\% confidence level.
\protect{\label{fig:S3}}}
	\end{center}
	\end{figure}

	\begin{figure}[h]
	\begin{center}
	\includegraphics[angle=0,width=0.8\textwidth]
 		{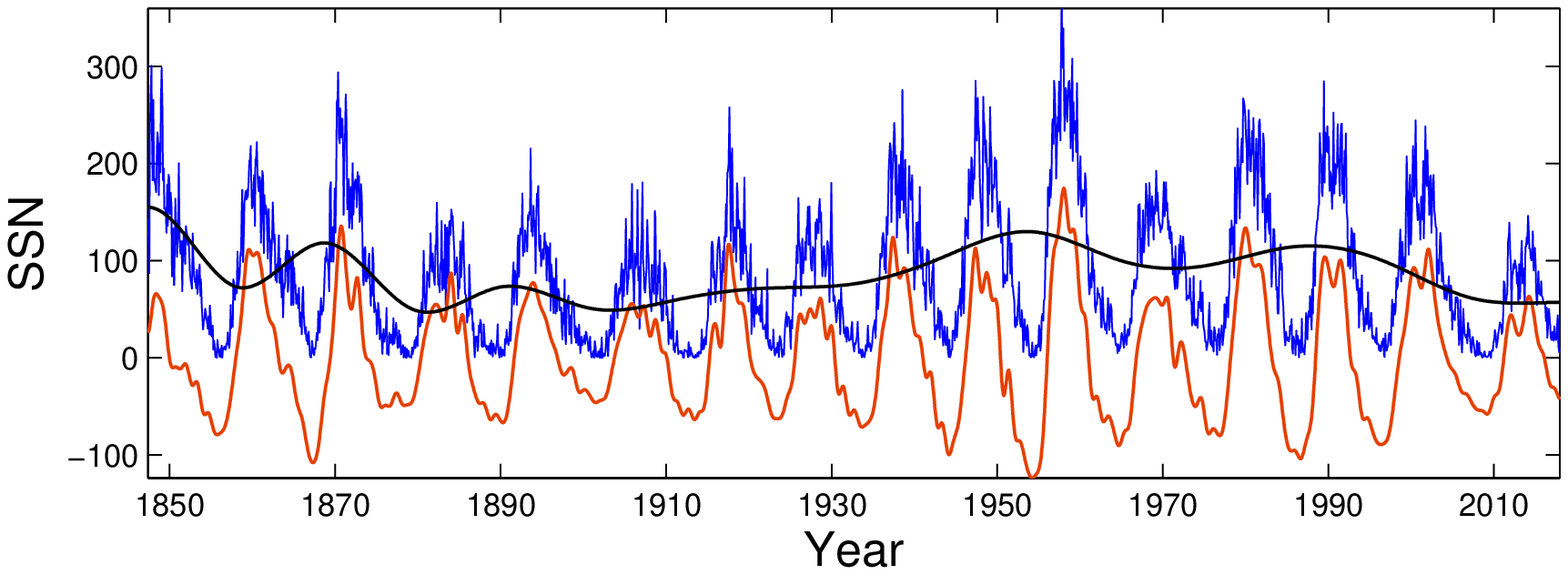}
	\caption{ 
	The original monthly SSN series (blue), lowpass-filtered series
(black), and bandpass-filtered series (red).
\protect{\label{fig:S4}}}
	\end{center}
	\end{figure}

	\begin{figure}[h]
	\begin{center}
	\includegraphics[angle=0,width=0.8\textwidth]
 		{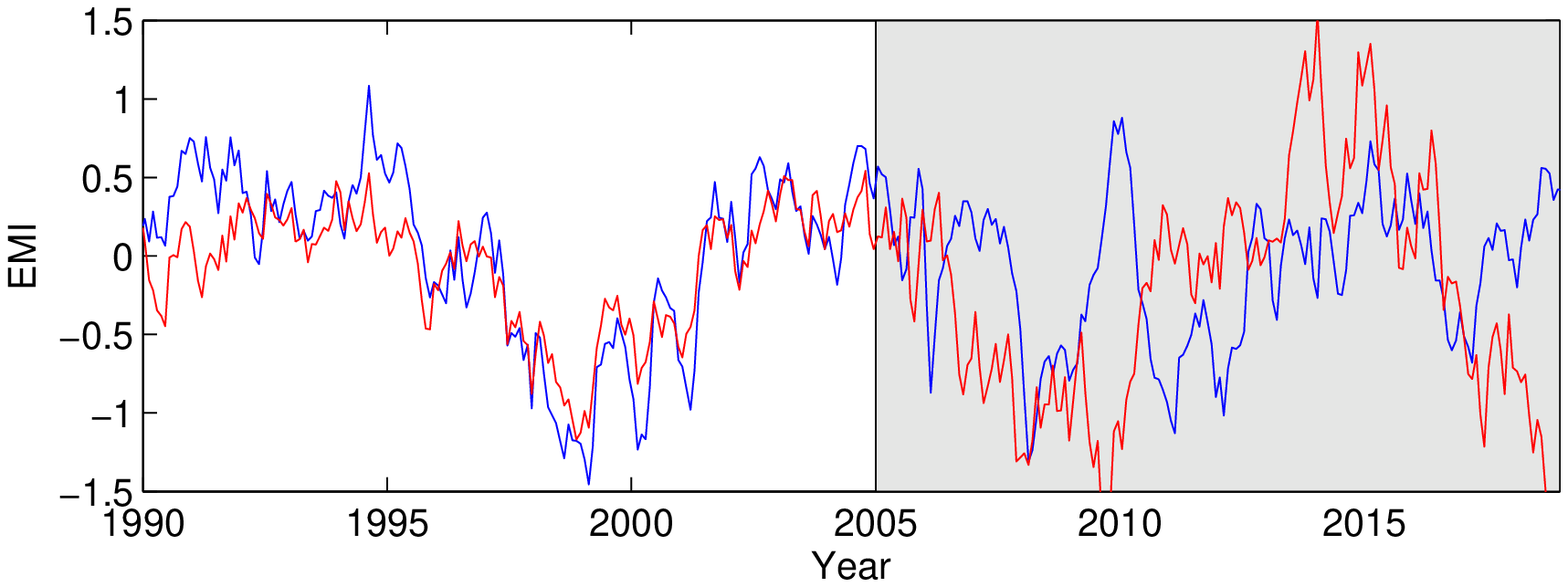}
	\includegraphics[angle=0,width=0.8\textwidth]
 		{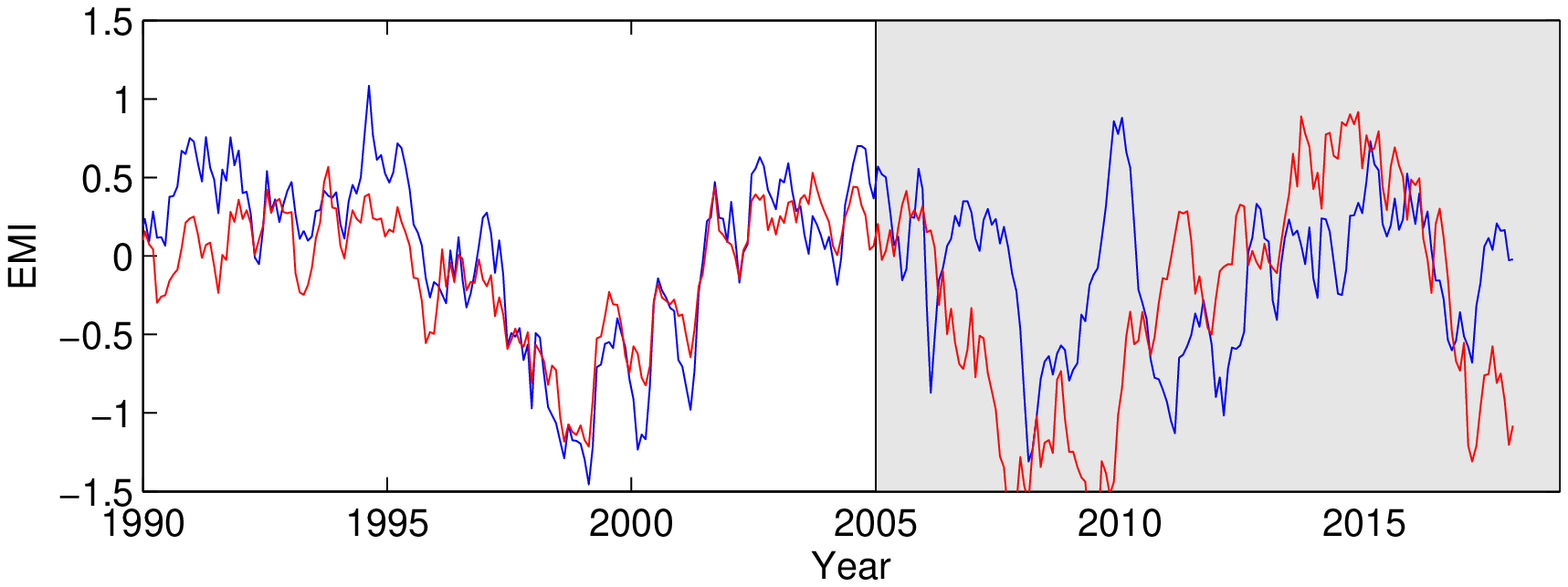}
	\caption{ 
Projection of the El Ni\~no Modoki Index (EMI) based solely on the SSN
22-50 years ago with a simple linear regression model. Upper: The SSN
series as input is not filtered. Lower: The SSN series is low-pass
filtered. The observed EMI is in blue, while the projected is in red.
\protect{\label{fig:S5}}}
	\end{center}
	\end{figure}

	\begin{figure}[h]
	\begin{center}
	\includegraphics[angle=0,width=0.8\textwidth]
 		{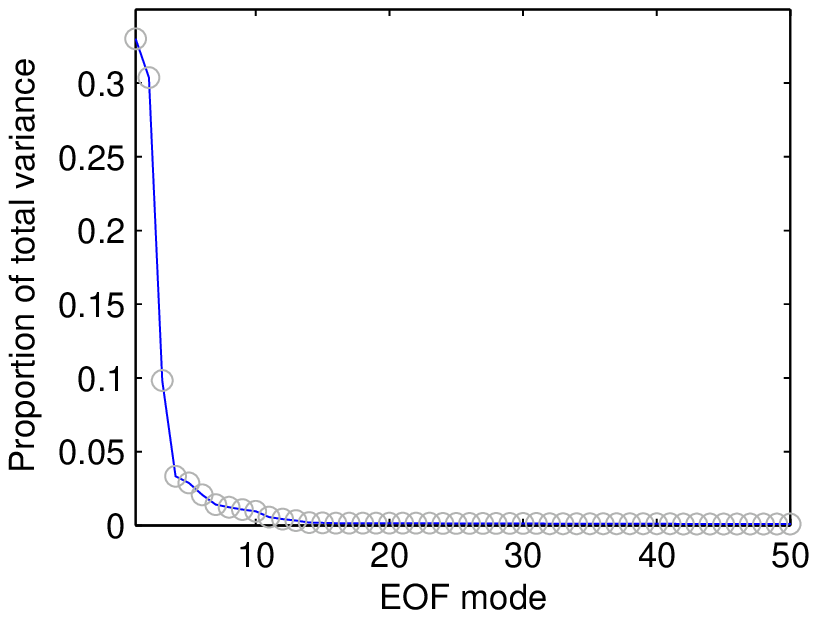}
	\caption{ NormaliEd variance vs. EOF mode. The EOF analysis is
performed with the SSN time series with delays from 22-50 years, which 
form a vector with 336 members of series.
\protect{\label{fig:S6}}}
	\end{center}

	\end{figure}
	\begin{figure}[h]
	\begin{center}
	\includegraphics[angle=0,width=1\textwidth]
 		{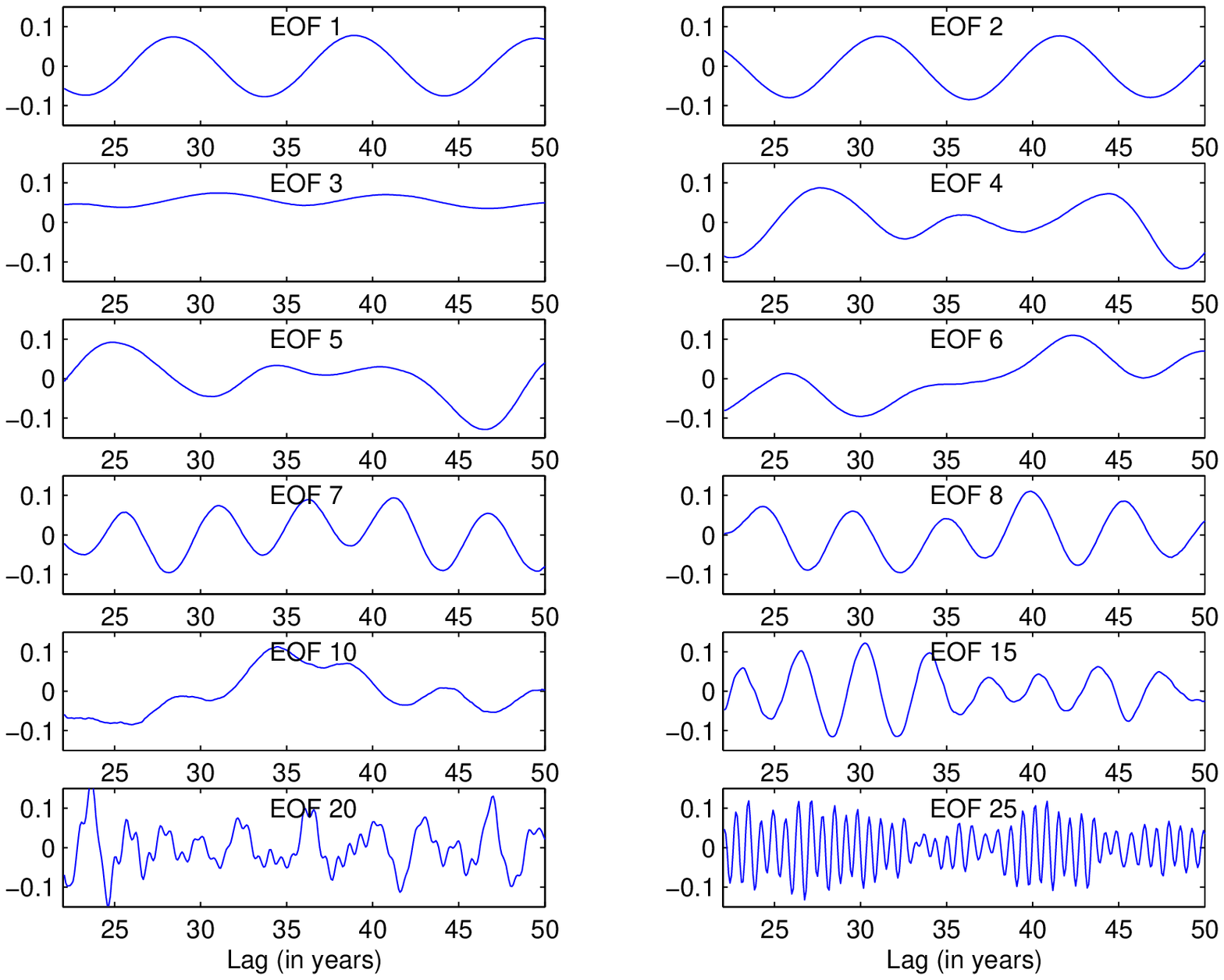}
	\caption{ Examples of the EOF modes as described in Supplementary
	Figure~\ref{fig:S6}.
\protect{\label{fig:S7}}}
	\end{center}
	\end{figure}

	\begin{figure}[h]
	\begin{center}
	\includegraphics[angle=0,width=0.8\textwidth]
 		{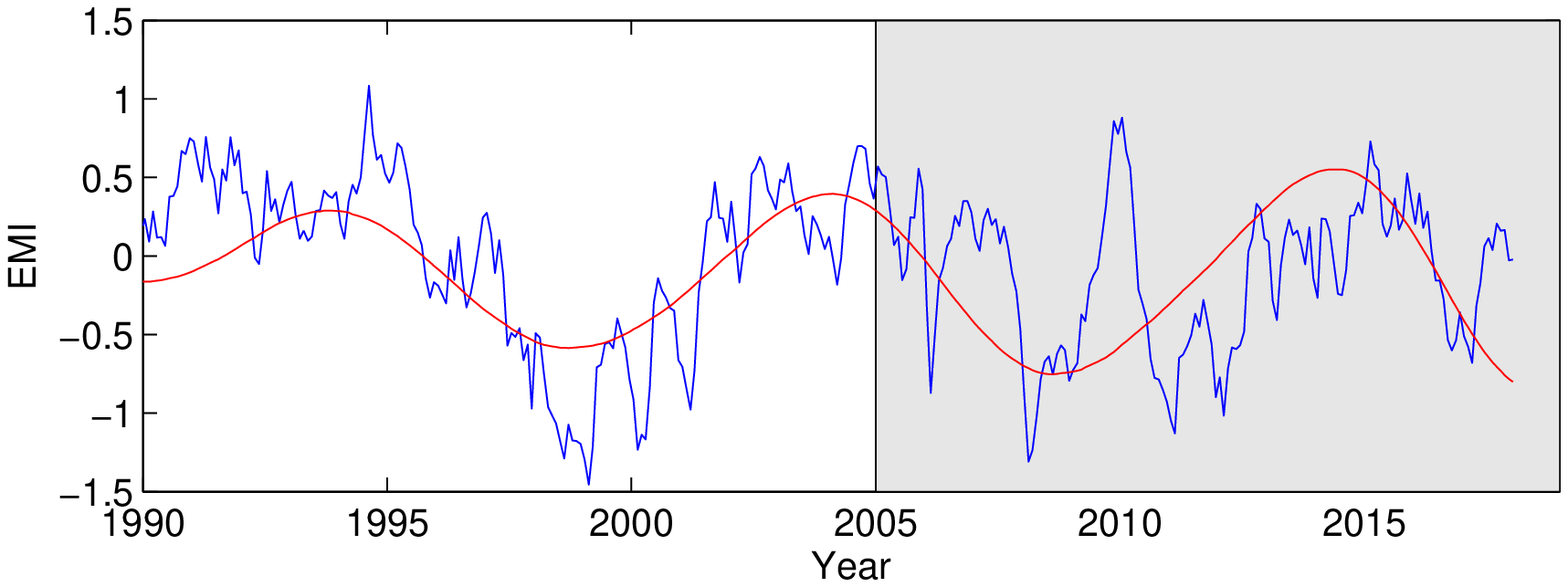}
	\includegraphics[angle=0,width=0.8\textwidth]
 		{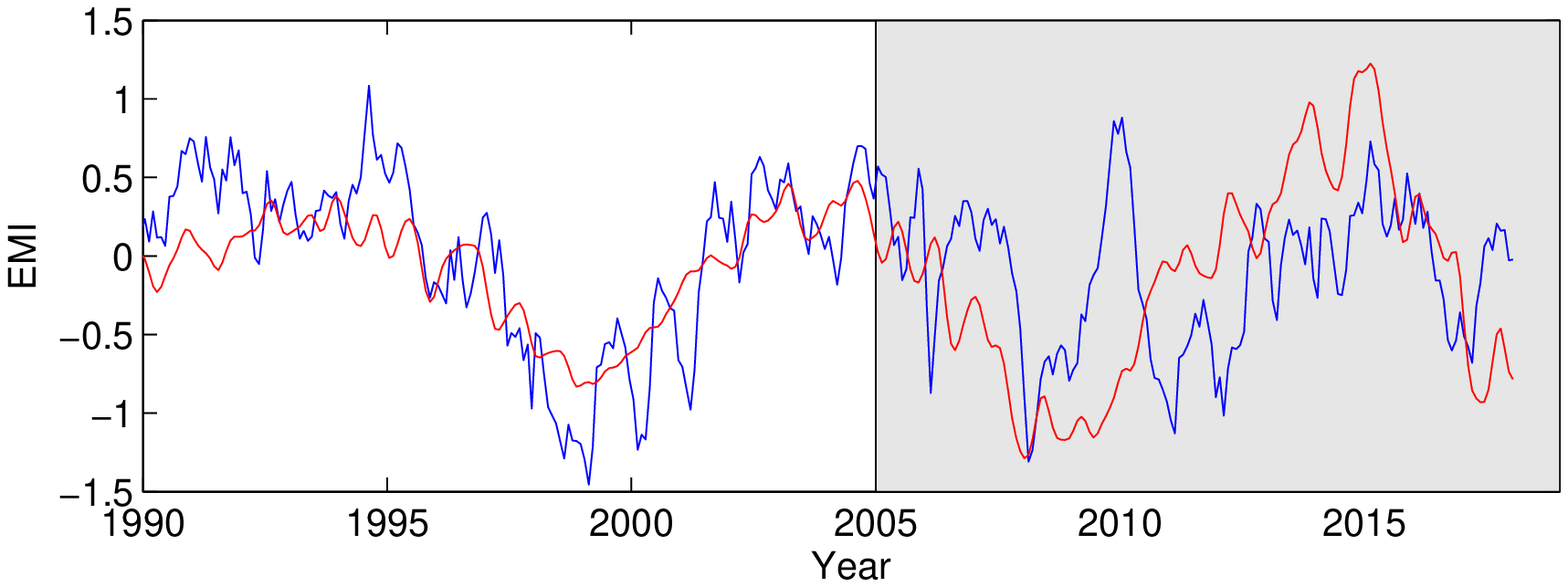}
	\caption{ As Figure 2b, but the projection uses, respectively, 
	the first 8 and 50 principal components as inputs.
		\protect{\label{fig:S8}}}
	\end{center}
	\end{figure}

\end{document}